\documentclass[11pt]{JHEP3}
\usepackage{epsfig,multicol,bbm}
\usepackage{cite}
\usepackage{amsmath,amsfonts,amssymb,longtable}
\usepackage{array} 
\usepackage{graphicx}
\usepackage{subfig}
\input epsf.sty

\newcommand{\I}{\mathrm{i}}
\newcommand{\E}[1]{\ensuremath{\mathrm{E}_{#1}}} 
\newcommand{\SU}[1]{\ensuremath{\mathrm{SU}(#1)}}
\newcommand{\SO}[1]{\ensuremath{\mathrm{SO}(#1)}}
\newcommand{\U}[1]{\ensuremath{\mathrm{U}(#1)}}
\newcommand{\Z}[1]{\ensuremath{\mathbb{Z}_{#1}}} 
\newcommand{\x}{\ensuremath{\times}}
\newcommand{\maN}{\ensuremath{\mathcal{N}} }
\newcommand{\bs}[1]{\ensuremath{\boldsymbol{#1}} }
\newcommand{\bsb}[1]{\ensuremath{\boldsymbol{\overline{#1} }}}
\DeclareMathOperator{\diag}{diag}

\usepackage{cancel, color}
\let\normalcolor\relax

\def\bra{\langle}
\def\ket{\rangle}
\def\beq{\begin{equation}}
\def\eeq{\end{equation}}
\newcommand{\C}[1]{\mathcal{#1}}
\def\ov{\overline}
\def\ap{\alpha^\prime}

\def\tr{\mathrm{tr}}
\def\nn{\nonumber}

\title{\huge Kinetic Mixing of U(1)s in Heterotic Orbifolds}

\author{Mark Goodsell \!$^1$, Sa\'ul Ramos-S\'anchez$^{2}$, Andreas Ringwald$^3$ \\ \\
${}^1$CERN, Theory Division, CH-1211 Geneva 23, Switzerland \\
${}^2$Department of Theoretical Physics, Physics Institute, UNAM, Mexico D.F. 04510, Mexico \\ 
${}^3$Deutsches Elektronen-Synchrotron DESY, Hamburg, Germany \\ \\
E-mail: {\email{mark.goodsell@desy.de}\,,
\email{ramos@fisica.unam.mx}\,, 
\email{andreas.ringwald@desy.de}}
}

\preprint{DESY 11-184\\
CERN-PH-TH/2011-253}

\abstract{
We study kinetic mixing between massless \U1 gauge symmetries in
the bosonic formulation of heterotic orbifold compactifications. 
For non-prime \Z{N} factorisable orbifolds, we find a simple expression of the
mixing in terms of the properties of the $\maN=2$ subsectors, which helps
understand under what conditions mixing can occur. With this tool, we 
analyse \Z{6}-II heterotic orbifolds and find non-vanishing mixing
even without including Wilson lines. We show that some
semi-realistic models of the Mini-Landscape admit supersymmetric vacua
with mixing between the hypercharge and an additional \U1, which can be broken
at low energies. We finally discuss some phenomenologically appealing
possibilities that hidden photons in heterotic orbifolds allow. 
}


\begin{document}

\section{Introduction}
\label{SECTION:INTRO}

\subsection{Motivation}

A good motivation for the existence of additional \U1 gauge symmetries
is the ``Dark Force" scenario. In this setting, dark matter arises from
a standard model (SM) singlet charged under a hidden \U1, whose gauge 
boson has a GeV mass and mixes kinetically with the observable 
photon~\cite{Feldman:2006wd,Feldman:2007wj,Pospelov:2007mp,Hooper:2008im,ArkaniHamed:2008qn,Pospelov:2008jd,ArkaniHamed:2008qp}.
Several features of such scenarios have been studied~\cite{Chun:2008by,Cheung:2009qd,Katz:2009qq},
including supersymmetry~\cite{Hooper:2008im,ArkaniHamed:2008qp,Morrissey:2009ur,Cohen:2010kn,Andreas:2011in}, in which 
some of the contemporary puzzles, such as the data of PAMELA and Fermi, could find an
explanation. 
Furthermore, the wealth of experiments~\cite{Pospelov:2008zw,Essig:2009nc,Bjorken:2009mm} 
(see e.g.~\cite{Jaeckel:2010ni,Redondo:2010dp,Andreas:2011xf} for an overview) 
capable of probing hidden \U1s
over a very wide range of hidden gauge-boson mass and kinetic mixing values also motivates
the study of the phenomenological potential of hidden sectors. 
On the other hand, string theory quite often offers a plethora of
Abelian symmetries which might well help to embed the Dark Force scenario
in an ultra-violet complete and globally consistent theory. 
Thus, it seems natural to ask to what extent this idea is consistent with string constructions.
As a starting point, we could  well ask what hidden-sector models are 
\emph{generic} and what the generic parameters for them are. In particular,
we can explore the amount of kinetic mixing that promising models exhibit.

Predictions for kinetic mixing and its phenomenology were considered in type II strings in 
\cite{Abel:2003ue,Lust:2003ky,Abel:2006qt,Abel:2008ai,Goodsell:2009xc,Gmeiner:2009fb,Goodsell:2009pi,Bullimore:2010aj,Cicoli:2011yh}; 
in~\cite{Goodsell:2009xc,Cicoli:2011yh} both masses and mixings were studied, and it was 
argued that the Dark Force scenario could be accommodated provided that there 
is additional sequestering. However, it is to heterotic orbifolds that we shall turn in this paper.
Orbifold models in the fermionic formulation were the corner of the heterotic string landscape where kinetic mixing 
was originally searched for~\cite{Dienes:1996zr,Goodsell:2010ie} (it was also
considered in heterotic models in the geometric regime in \cite{Lukas:1999nh,Blumenhagen:2005ga}),
although the existence of interesting models (i.e. with non-zero mixing) was not established. 
We intend to resolve this issue, and also clarify under what conditions mixing may occur. To do so, we study 
kinetic mixing in heterotic orbifold models in the bosonic formulation and focus 
particularly on those of the \Z6--II orbifold
Mini-Landscape~\cite{Lebedev:2006kn,Lebedev:2008un}, which 
are known to display many 
phenomenologically appealing features~\cite{Nilles:2008gq}.
Indeed, we hope that this paper paves the way for more exploration of the rich phenomenology 
of kinetically mixed hidden sectors possible in heterotic models.

\subsection{Kinetic mixing in supersymmetric theories}

In an unbroken supersymmetric theory, there is only one possible operator that can yield kinetic mixing. 
It appears  in the gauge kinetic part of
the supergravity Lagrangian, and is thus a holomorphic function of other fields: 
\begin{equation}
\mathcal{L} \supset \int d^2 \theta  \left\{
\frac{1}{4 (g_a^h)^2} W_a W_a + \frac{1}{4(g_b^h)^2} W_b W_b
- \frac{1}{2}\chi_{ab}^h W_a W_b  \right\},
\end{equation}
where $W_a, W_b$ are the field strength superfields for the two U(1) gauge fields and $\chi_{ab}^h,g_a^h, g_b^h$ 
are the holomorphic kinetic mixing parameter and gauge couplings that must run only at one loop. In the canonical basis, 
\beq
\C{L}_{\mathrm{canonical}} \supset \int d^2 \theta \left\{
\frac{1}{4} W_a W_a + \frac{1}{4} W_b W_b
- \frac{1}{2}\chi_{ab} W_a W_b  \right\}\,,
\eeq
where the canonical kinetic mixing is given in terms of the holomorphic quantity 
by the Kaplunovsky-Louis type relation~\cite{Kaplunovsky:1994fg,Benakli:2009mk,Goodsell:2009xc}:
\beq
\frac{\chi_{ab}}{g_a g_b} = \Re(\chi_{ab}^h) + \frac{1}{8\pi^2} \mathrm{tr}\bigg( Q_a Q_b \log Z\bigg) - \frac{1}{16\pi^2} \sum_r n_r Q_a Q_b (r)\kappa^2 K.
\label{KLTypeFormula}\eeq
Here, $K$ is the full K\"ahler potential of the theory, and $Z$ is the K\"ahler metric of matter 
fields formed by the second derivative of $K$. $Q_{a,b}$ are the charge operators of the two \U1s.

The mixing can be read off from a one-loop string calculation of the S-matrix element for two gauge bosons.
At one loop this is identical to the 1PI diagram, and so we obtain the physical mixing. The 
calculation is analogous to that of gauge threshold corrections, 
and yields~\cite{Dienes:1996zr}
\beq
\frac{\chi_{ab}}{g_a g_b} = \frac{b_{ab}}{16\pi^2} \log \frac{M^2_S}{\mu^2} + \Delta_{ab}\,,
\label{eq:chirun}
\eeq
where $M_S$ is the string scale, and $b_{ab}$ and $\triangle_{ab}$ are mixed $\beta$ function 
coefficients and string threshold contributions, respectively.

The previous results have already been applied in several extensions of the SM, including string theory.
For example, in~\cite{Dienes:1996zr} it was 
found that none of their models satisfied $b_{ab}=0$, condition that 
ensures no running of the kinetic mixing below the string scale and corresponds to 
eliminating (chiral) light states charged under both \U1s. They also pointed out that the phenomenological 
problems associated with these states can be dodged once they are lifted out of 
the spectrum after
certain SM singlets attain vacuum expectation values (VEVs) (although this would 
be very challenging to calculate in practise). Therefore, in their work $\Delta_{ab}$ 
was considered of prime interest as an unambiguous indicator of 
kinetic mixing. We shall take the same approach: while it would be particularly pleasing to obtain models 
for which $b_{ab} =0$, we shall not restrict ourselves to that case. Unfortunately, $\Delta_{ab}$ was 
calculated for various models then available and exactly zero was found in each case. This is in contrast 
to \cite{Kaplunovsky:1995jw}, which contained examples of \Z2\x\Z2
orbifolds without Wilson lines for which $b_{ab}= 0$ and $\Delta_{ab} \ne 0$ but did not emphasize
this result; however, the \U1s descended from a higher-dimensional \SU2 in each case.

Our discussion in the present work is structured as follows.
In section \ref{SECTION:KM}, we address the general computation of kinetic mixing
in heterotic orbifolds, and attempt to 
obtain a better understanding of when mixing is possible. Then in section~\ref{SECTION:EX}, we
present examples with and without Wilson lines where $\Delta_{ab}$ is non-zero and briefly
study their phenomenology. We show that the models of the \Z6--II heterotic Mini-Landscape
can accommodate mixing $\Delta_{ab} \ne 0$ between the hypercharge and additional hidden forces in 
MSSM-like supersymmetry-preserving vacua. 
In section~\ref{sec:phenohiddenphotons}, we briefly explore how hidden \U1
symmetries emerging from heterotic orbifolds could yield a Dark Force scenario and other 
interesting phenomenology. Finally, 
in section~\ref{sec:discussion}, we discuss our results and provide an outlook. The appendices
are devoted to the details of some sample models.

A final remark is in order. In this work, we do \emph{not} consider the case 
where $b_{ab}=0$ and $\Delta_{ab}=0$ simultaneously. However, we note that in 
these cases, mixing can still be generated by splitting the masses of multiplets charged under both \U1s; 
in the case of light fields obtaining VEVs, this can generate sizable mixing, while splitting 
due to supersymmetry breaking seems to generate only an incredibly small effect~\cite{Goodsell:2009xc,Goodsell:2010ie}.

\section{Kinetic mixing in heterotic orbifolds}
\label{SECTION:KM}

\subsection{Heterotic orbifolds}

Let us start by introducing the main concepts that are important for computing the kinetic mixing
between \U1s in the context of heterotic orbifolds 
(for reviews of these constructions, see~\cite{Bailin:1999nk,Choi:2006qh,RamosSanchez:2008tn,Vaudrevange:2008sm}).
The \Z{N} orbifolds in which we are interested are defined as quotients of factorisable
six-dimensional tori $T^6=T^2\x T^2\x T^2$ divided by a discrete set of isometries of
$T^6$ that form the so-called point group. In \Z{N} heterotic orbifolds, 
the point group contains all different powers of a discrete rotation generator 
$\vartheta = \diag(e^{2\pi\I v_1},e^{2\pi\I v_2},e^{2\pi\I v_3})$,
where each entry of the {\it twist vector} $v$ encodes the action of the orbifold on 
each 2-torus and $v_1+v_2+v_3=0$. Different powers $k$ of $\vartheta$ ($0\leq k < N$) define
different {\it twisted sectors}. If the action of $\vartheta^k$ is trivial on only 
one $T^2$, i.e. if $kv_i =0\mod 1$ for an $i$, $kv$ defines an $\maN=2$ subsector.
As we shall see below, orbifold constructions that may allow for kinetic 
mixing require the existence of these $\maN=2$ subsectors. Consequently, by inspecting
all admissible twist vectors (complete lists can be found in~\cite{Kobayashi:1991rp}),
one finds that all \Z{N} orbifolds with non-prime $N$ are candidates to provide kinetic 
mixing.\footnote{We note that also \Z{N}\x\Z{M} orbifolds fall in this category, but we will not
consider them here.}

We decompose \U1 gauge fields in terms of the Cartan generators of the initial 
\E8\x\E8 (or -- but not here -- $\mathrm{Spin}(32)/\mathbb{Z}_2$), $H_I (I=1,\ldots,16)$, as 
\begin{align}
Q_a =& \sum_I t_a^I H_I.
\label{eq:defQt}
\end{align}
Then, a state with left-moving momenta $p_I$ will have \U1 charge $t_a^I p_I$.
The vector $t_a$  will be frequently called the generator of the $\U1_a$ symmetry.
The operators on the worldsheet corresponding to the gauge bosons contain holomorphic currents $j(z)$ satisfying the OPE
\begin{align}
j^I (z) j^J (0) \sim \frac{k^I \delta^{IJ}}{z^2} + \frac{\I f^{IJ}_{\ \ K}}{z} j^K(0)\,,
\end{align} 
where $k^I$ is the level of the algebra, which we shall take to be normalised to 2. 
For our \U1s, the currents 
appear in the combinations $\sum_I t_a^I j^I(z)$ and so the tree-level gauge kinetic function satisfies 
\begin{align}
f_{ab} =& S \sum_I k^I t_a^I t_b^I\,,
\end{align}
where $S$ is the (bosonic part of) the dilaton/axion chiral superfield.
Hence the independent \U1s satisfy $\sum_I t_a^I t_b^I = \delta_{ab} $ and have gauge kinetic function 
$f_{ab} = 2S \delta_{ab}$. Note that \emph{this is not the same as $\tr(Q_a Q_b) = \delta_{ab} $}. Due to 
this orthogonality, there is no $\C{O}(z^{-2})$ term in the OPE of two different \U1s.

We are interested in the mixing between non-anomalous \U1s, i.e. between Abelian 
symmetries satisfying the universality condition~\cite{Casas:1987us,Kobayashi:1996pb}
\beq
\frac{1}{2|t_b|^2}\tr(Q_b^2 Q_a) = \frac1{6|t_a|^2}\tr\, Q_a^3 =\frac1{24}\tr\, Q_a=0\quad \text{for all } a,b\,,
\label{eq:universality}
\eeq
where the traces run over all chiral-matter fields.\footnote{Although we will mostly take $|t_a|^2=1$, as we 
mentioned before, we shall use the GUT-compatible hypercharge normalisation $|t_Y|^2=5/6$.}
It is well-known that the orbifold limit of the heterotic string has commonly (at most)
one anomalous symmetry $\U1_\mathrm{anom}$~\cite{Witten:1984dg}. This anomaly is cancelled 
via the Green-Schwarz mechanism~\cite{Green:1984sg} which implies that, for the anomalous 
case,~\eqref{eq:universality} takes the form
\beq
\frac{1}{2|t_b|^2}\tr(Q_b^2 Q_\mathrm{anom}) = \frac1{6|t_\mathrm{anom}|^2}\tr\, Q_\mathrm{anom}^3 
   =\frac1{24}\tr\, Q_\mathrm{anom}= 8\pi^2 \delta_{GS}\quad \text{for all } b\,,
\label{eq:universalityAnom}
\eeq
where  $\delta_{GS}$ is the universal Green-Schwarz coefficient.
The existence of $\U1_\mathrm{anom}$ induces the Fayet-Iliopoulos $D$-term~\cite{Dine:1987xk} 
given (in string units) by
\beq
\xi = g_s^2 \,\delta_{GS} = \frac{g_s^2 \,\tr\, Q_\mathrm{anom}}{192\pi^2}\,,
\label{eq:FIterm}
\eeq
where $g_s$ is the string coupling. It is precisely the appearance of this term what renders
this anomalous symmetry harmless for phenomenology. Since we do not expect supersymmetry to
be broken at the compactification scale in realistic models, in an acceptable orbifold vacuum 
$\xi$ must be cancelled. This means that certain matter fields charged under $\U1_\mathrm{anom}$
need develop vacuum expectation values, breaking thereby this anomalous symmetry and avoiding
dangerous mixing effects. Therefore, for practical purposes, we shall ignore the anomalous
\U1 and focus on constructions with two or more non-anomalous \U1s.

Once the twist vector is set, different \Z{N} orbifold models arise from the different gauge 
embeddings that satisfy the modular invariance conditions~\cite{Senda:1987pf, Dixon:1986jc, Vafa:1986wx, Ploger:2007iq}
which ensure an effective theory free of any anomalies. 
We are then interested in admissible shift vectors $V$ and Wilson lines ${\cal A}_i$ 
that comply with (no summation implied)
\begin{subequations}\label{eq:ModInv}
\begin{eqnarray}
\label{eq:ModInvV}
N(V^2-v^2)&=&0\mod 2\,.\\
  \label{eq:ModInvVAa}
  N_i\,{\cal A}_{i}\cdot V  & = & 0 \mod 2\,, \\
  \label{eq:ModInvAaAa}
  N_i\,{\cal A}_{i}^2  & = & 0 \mod 2\,, \\
  \label{eq:ModInvAaAb}
  \gcd(N_{i},N_{j})\,{\cal A}_{i}\cdot {\cal A}_{j}  & = & 0 \mod 2 \qquad (i\neq j)\,,
\end{eqnarray}
\end{subequations}
where  the order of the Wilson lines $N_i$ is constrained by the geometry of $T^6$
and $\gcd$ stands for the greatest common denominator. We focus here on the ten-dimensional
heterotic string theory with gauge group \E8\x\E8, which is broken down to the four-dimensional
group by the action of the shift and Wilson lines.

\subsection{Kinetic mixing}

For these models, the kinetic mixing at the string threshold is given by
\begin{align}
\Delta_{ab} \equiv& \frac1{16\pi^2} \int \frac{d^2 \tau}{\mathrm{Im} \tau} [B_{ab} (\tau) - b_{ab}]  \nn\\
\text{with }\quad B_{ab} (\tau) \equiv& - \mathrm{Str}( \ov{Q}_H^2 Q_a Q_b e^{\ap M_R^2 2\pi i \ov{\tau}} e^{\ap M_L^2 2\pi i \tau})\,,
\end{align}
where $M_R, M_L$ are the masses of respectively right and left moving states in the theory
(which are constrained to be equal), $\ov{Q}_H$ denotes the helicity operator, and
\beq
b_{ab} \equiv - \mathrm{Str}_{\rm massless} ( \ov{Q}_H^2 Q_a Q_b) 
\eeq 
with $\mathrm{Str}_{\rm massless}$ being a supertrace over massless states.

Note that the result is entirely analogous to gauge threshold corrections, except 
there is no moduli-independent piece proportional to the level: this is easily seen 
as being due to the lack of an $\C{O}(z^{-2})$ term in the OPE of the currents. 
Following the reasoning in~\cite{Dixon:1990pc,Kaplunovsky:1995jw},
the result can be expressed as follows
\begin{align}
\label{eq:generalkineticmixing}
\Delta_{ab} =&  \sum_{i} \frac{b_{ab}^i |G^i|}{16\pi^2 |G|} \bigg[ \log \bigg(|\eta(T_i)|^4 \mathrm{Im} (T_i)\bigg) 
                 + \log \bigg(|\eta(U_i)|^4 \mathrm{Im} (U_i)\bigg) \bigg]\,,
\end{align}
where the sum runs over all order $|G^i|$ $\maN=2$ subsectors of the orbifold of order $|G|$, $T_i, U_i$ 
are the untwisted K\"ahler and complex structure moduli (in Planck units) corresponding to the torus fixed in the subsector 
$i$, and the $\beta$-like mixing coefficient associated to the $i$-th fixed torus $b_{ab}^i$ is given by
\begin{align}
\label{eq:babN=2}
b_{ab}^i \equiv& - \frac{11}{6} \tr^i_V (Q_a Q_b) + \frac{1}{3} \tr^i_F (Q_a Q_b) +  \frac{1}{12} \tr^i_S (Q_a Q_b) \nn\\
=&\frac{1}{4} \bigg[- 6 \tr^i_V (Q_a Q_b)  +  \tr^i_S (Q_a Q_b) \bigg].
\end{align}
Here, the traces in the first line are over the vectors, Weyl fermions and real scalars respectively, and in 
the second line we have simplified the expressions using supersymmetry -- this is the form used in a brute force 
calculation of the mixing, by summing over all bosonic states in the appropriate sector of the theory, identifying 
them as vectors or scalars, and weighting accordingly with $\sum_{I,J} (t_a^I p_I) (t_b^J p_J)$. Of course, the sum over states
is equivalent to the more familiar sum over $\maN=2$ vector and hypermultiplets 
$$b_{ab}^i = \frac{1}{2} \left( -2 \tr^i_{V, \maN=2} (Q_a Q_b) + \tr^i_{H,\maN=2} (Q_a Q_b)\right)\,, $$
where $Q_{a,b}$ are defined in eq.~\eqref{eq:defQt}, 
i.e. they correspond to the $\maN=1$ Abelian generators, whereas all summations in eq.~\eqref{eq:babN=2} 
are over $\maN=2$ states.

We would also like now to understand when we can obtain kinetic mixing, and provide a simple formula to 
explain its presence or absence. To this end, we shall rewrite $b_{ab}^i$ in terms of the properties of
the gauge subgroups $b'$ in the different $\maN=2$ sectors~\footnote{The indices $a,b,\ldots$ denote $\maN=1$
gauge subgroups and $a',b',\ldots$ refer to the $\maN=2$ theories.}. In doing so, it is important to trace
the origin of the \U1s in the $\maN=2$ sectors.
For example, there may be \U1s in an $\maN=2$ sector that may be broken or not in 
the full orbifold, and the $\maN=1$ \U1s may also arise from the non-Abelian groups in the $\maN=2$ sector. 
The simple roots of the $\maN=2$ non-Abelian subgroups $b'$ shall be denoted by 
$\hat{\alpha}^{b'}_i$ with $i$ running from $1$ to the sum of the ranks of the non-Abelian groups $r$;
the $\maN=2$ Abelian generators are denoted by $t_{b'}$ (clearly, the sum of $\hat\alpha^{b'}_i\,$s and $t_{b'}\,$s is 16). 
For non-Abelian groups, the $\hat{\alpha}_i^{b'}$ 
are normalised to $\bra \hat{\alpha}_i^{b'} \hat{\alpha}_j^{c'} \ket = \delta^{b'c'} C_{ij}^{b'}$, where $C^{b'}$
is the Cartan matrix of the group $b'$. As for $\maN=1$, in the Abelian case we have
$\bra t_{b'} t_{c'} \ket = \delta_{b'c'}$. We can then write the
$\maN=1$ $\U1_a$ generators of interest $t_a$ as linear combinations:
\begin{align}
\label{eq:Neq2combination}
t_a =& \sum_{i=1}^{r} m^{b',i}_a \hat{\alpha}^{b'}_i + \sum_{b'=r+1}^{16}n^{b'}_a t_{b'}\,,\,\qquad m_a^{b',i},n_a^{b'}\in\mathbb{R}\,.
\end{align}
Defining additionally the matrix $b_{\U1}^{\maN=2}$ as
\begin{equation}
\left(b_{\U1}^{\maN=2}\right)_{a'b'} = \frac{1}{2} \left( -2 \tr_{V, \maN=2} (Q_{a'} Q_{b'}) + \tr_{H,\maN=2} (Q_{a'} Q_{b'})\right)
\end{equation}
(with $Q_{a',b'}$ being the the analog of~\eqref{eq:defQt} for $\maN=2$),
we are ready to rewrite the mixing coefficient $b_{ab}^i$
in terms of the previous matrix, the $\maN=2$ $\beta$-function coefficients $b_{b'}^{\maN=2}$ of
the non-Abelian groups, and the Cartan matrices $C^{b'}$:
\begin{align}
\label{eq:mixingsimple}
b_{ab}^i =& \left(m^{b'}_a C^{b'} m^{b'}_b\right) b_{b'}^{\maN=2} + 2 \left(n_a\, b_{\U1}^{\maN=2}\, n_b\right)\,.
\end{align} 

This provides an explanation for when kinetic mixing can be present: the \U1s must either 
contain overlapping components in non-Abelian gauge groups in the $\maN=2$ sector, or must 
derive from \U1s that mix at that level. 
Note that clearly if the generators of the
\U1s lie entirely in separate \E8s, then only the second possibility is available provided
that the off-diagonal entries of $b_{\U1}^{\maN=2}$ are non-trivial. 

A final remark is in order. The actual value of kinetic mixing at low energies depends
not only on the high energy contribution from the $\maN=2$ subsectors, $b_{ab}^i$ (or $\Delta_{ab}$), but 
also importantly on the $\maN=1$ $\beta$ function coefficient $b_{ab}$, as in eq.~\eqref{eq:chirun}.
Since this dependence might alter drastically any result coming from the high-energy string states, and wishing to have a hidden \U1, 
 it is desirable to take $b_{ab}=0$. In heterotic orbifolds, this requires the spectrum to be vector-like w.r.t. the \U1s, 
which imposes a strong constraint on the models. This can be contrasted to D-brane models, where this constraint can be 
readily satisfied by separating the branes supporting the \U1s, but it is then necessary to ensure the absence of 
\emph{mass mixing} due to the St\"uckelberg mechanism (even though the \U1s may not be anomalous) while 
preserving kinetic mixing \cite{Abel:2008ai}. This does not occur in the heterotic orbifold 
case since there are no axions that could generate the masses\footnote{In the geometric regime 
(or equivalently when the orbifold is blown up)~\cite{Lukas:1999nh,Blumenhagen:2005ga} it does happen.}.

\section{Kinetic mixing in \Z{N} heterotic orbifolds}
\label{SECTION:EX}

Since only non-prime \Z{N} orbifolds can produce non-trivial kinetic mixing, we shall
consider below one promising candidate: \Z6--II, which has been found to lead to plenty 
of models possessing many phenomenologically appealing features~\cite{Nilles:2008gq}. Therefore, 
an interesting question is whether explicit models of this type exhibit kinetic mixing. Remarkably, 
even \Z6--II orbifold models without Wilson lines display non-trivial values of $\Delta_{ab}$, 
as we now discuss.

\subsection{\Z6--II orbifolds without Wilson lines}

The \Z6--II orbifold is defined by the twist vector $v=1/6(1,2,-3)$ acting on the $T^6$ torus spanned by
the root lattice of $G_2\x\SU3\x\SO4$. The structure of these constructions allows only for three K\"ahler moduli
$T_1,T_2,T_3$ and a complex-structure modulus $U_3$, where the subindexes refer to the three 2-tori of $T^6$.  
Without Wilson lines, there are 61 different gauge embeddings $V$ that fulfill~\eqref{eq:ModInvV}, but 
only roughly $1/3$ of them may lead to kinetic mixing because they have two or more non-anomalous \U1s. 

The stringy contributions to the kinetic mixing come from the
two $\maN=2$ subsectors of \Z6--II orbifolds: a \Z3 subsector 
generated by $2v$ that comprises the $k=2,4$ twisted sectors and leaves the third $T^2$ invariant, and 
a \Z2 subsector generated by $3v$ which includes only the $\vartheta^3$ sector and leaves the second $T^2$
untouched. Therefore, we have
\begin{eqnarray}
\label{eq:DeltaZ6II}
\Delta_{ab}&=& \frac1{16\pi^2}\left\{  \frac{b_{ab}^2}{3} \log \bigg(|\eta(T_2)|^4 \mathrm{Im}(T_2)\bigg)\right. \nn\\
             &+& \left.\frac{b_{ab}^3}{2}\bigg[ \log \bigg(|\eta(T_3)|^4 \mathrm{Im}(T_3)\bigg)  + \log \bigg(|\eta(U_3)|^4 \mathrm{Im}(U_3)\bigg) \bigg]\right\}\,.
\end{eqnarray}

Using the methods described in section~\ref{SECTION:KM}, we can compute $b_{ab}^i$ for all
\Z6--II orbifolds without Wilson lines. We find that there are 10 models with 
\beq
\label{eq:resultZ6IInoWL}
3 \lesssim |b_{ab}^i| \lesssim 90 \qquad \Rightarrow \qquad 10^{-2} \lesssim |\Delta_{ab}|\lesssim 10^{-1}\,,
\eeq
where we have used moduli of order unity to estimate the values of $\Delta_{ab}$. Remarkably, even in this
simple scenario, there is kinetic mixing.

\paragraph{An example.} Consider now the model defined by the shift vector
\beq
\label{eq:VmodelnoWL}
V~=~\frac1{12}(10,2,2,2,0,0,0,0)(9,1,1,1,1,1,1,1)\,.
\eeq
The unbroken gauge group is then $\SO8\x\SU4\x\SU7\x\U1_\mathrm{anom}\x\U1_1\x\U1_2$.
The (coefficients of the Cartan-expansion of the) \U1 generators are given by
\begin{subequations}\label{eq:tsmodelnoWL}
\begin{eqnarray}
t_\mathrm{anom}&=&\tfrac1{4\sqrt2}(0,0,0,0,0,0,0,0)(5,1,1,1,1,1,1,1)\,,\\
t_1 &=& \tfrac1{4\sqrt{14}}(0,0,0,0,0,0,0,0)(-7,5,5,5,5,5,5,5)\,,\\
t_2 &=& \tfrac1{2}(-1,1,1,1,0,0,0,0)(0,0,0,0,0,0,0,0)\,,
\end{eqnarray}
\end{subequations}
(where we distinguish between the components of the first and second \E8s)
which satisfy the orthogonality constraint $t_a\cdot t_b=\delta_{ab}$. An interesting
feature of this model is that the relevant \U1s stem from different \E8 groups;
however, the $\maN=1$ matter spectrum leads to $\tr(Q_1Q_2)=-\frac{38\sqrt{14}}{15}\neq0$, 
implying that, although there is no kinetic mixing at tree level, at one-loop kinetic
mixing may appear. In fact, in this case evaluating~\eqref{eq:DeltaZ6II} results in
\begin{eqnarray}
\label{eq:DeltamodelnoWL}
\Delta_{12}&=&\frac{b_{12}^3}{2\x16\pi^2 }\bigg[ \log \bigg(|\eta(T_3)|^4 \mathrm{Im}(T_3)\bigg)  + \log \bigg(|\eta(U_3)|^4 \mathrm{Im}(U_3)\bigg) \bigg]\nn\\
           &=&-\frac{6\sqrt{14}}{16\pi^2}\bigg[ \log \bigg(|\eta(T_3)|^4 \mathrm{Im}(T_3)\bigg)  + \log \bigg(|\eta(U_3)|^4 \mathrm{Im}(U_3)\bigg) \bigg]\,,
\end{eqnarray}
which is about $10^{-1}$ for order-one moduli.

This result is interesting because it shows that, similarly to what happens in type II 
scenarios with D3-branes at \Z6--II orbifold singularities~\cite{Bullimore:2010aj}, \Z6--II
orbifold compactifications of the heterotic string allow for kinetic mixing. The advantage of the latter is
that the \U1s are not located at singularities and, therefore, it is not necessary to 
build by hand a suitable pair of \U1s that leads to this outcome. A shortcoming
of the model presented here is of course that it has no chance of being a description
of our universe, since it does not even exhibit the gauge group of the SM.

\subsection{Semi-realistic \Z6--II orbifolds with kinetic mixing}
\label{sec:semirealistic}

Introducing discrete Wilson lines in \Z6--II orbifolds leads to a large class of
semi-realistic models with an observable sector displaying the exact spectrum of the MSSM 
and other phenomenologically desirable properties~\cite{Nilles:2008gq}. In the hidden
sector, there are typically some Abelian gauge symmetries, which can be broken 
spontaneously in explicit supersymmetric vacua~\cite{Lebedev:2007hv,Brummer:2010fr}.
However, also supersymmetric vacua leading to two or more additional massless
\U1s exist~\cite{Lebedev:2009ag}, which may lead to observable kinetic mixing. 
In this section, we explore this possibility.

Before computing the kinetic mixing, a second effect of the presence of Wilson line
backgrounds must be considered. It is known that the original modular symmetry $SL(2,\Z{})$
is typically broken to its congruence subgroups $\Gamma_n(N),\Gamma^n(N)$  
by the Wilson lines~\cite{Spalinski:1991vd,Erler:1991ju,Bailin:1993ri}
(for some $n,N$ that depend on the chosen Wilson lines).
As explained e.g. in~\cite[\S 3.1]{Parameswaran:2010ec}, this breakdown forces the modular functions
$\eta(T_i),\eta(U_i)$ to be replaced by $\eta(p_i\,T_i),\eta(q_i\,U_i)$, where the factors $p,q$ depend
on the resulting modular subgroup.\footnote{E.g. $p,q=N$ for the subgroup $\Gamma_0(N)$.}
Therefore, including the effect of Wilson lines~\eqref{eq:DeltaZ6II} takes the form
\begin{eqnarray}
\label{eq:DeltaZ6IIWL}
\Delta_{YX} &=& \frac1{16\pi^2}\left\{ \frac{b_{YX}^2}{3} \log \bigg(|\eta(p_2\,T_2)|^4 \mathrm{Im}(T_2)\bigg)\right.\nn\\
            &+& \left.\frac{b_{YX}^3}{2}\bigg[ \log \bigg(|\eta(p_3\,T_3)|^4 \mathrm{Im}(T_3)\bigg)  + \log \bigg(|\eta(q_3\,U_3)|^4 \mathrm{Im}(U_3)\bigg) \bigg]\right\}\,.
\end{eqnarray}

Let us study now the subset of semi-realistic 
\Z6-II orbifold models obtained in the Mini-Landscape~\cite{Lebedev:2006kn,Lebedev:2008un}.
We find that almost all models (255)
allow for mixing between the hypercharge and one or more additional \U1 symmetries. 
The stringy contribution to $\chi$ in these constructions is approximately
\beq
\label{eq:resultZ6II2WL}
0.1 \lesssim |b_{YX}^i| \lesssim 7 \qquad \Rightarrow \qquad 10^{-4} \lesssim |\Delta_{YX}|\lesssim 10^{-2}\,,
\eeq
where, as before, moduli are assumed to be order one. (See appendix~\ref{app:moremodels}, for the details 
of some sample models of this kind.) This result contrasts with the one obtained
in semi-realistic orbifolds in the fermionic formulation, where $\Delta_{YX}$ was found to 
vanish~\cite{Dienes:1996zr}. 

It is more challenging to find possibly realistic vacua in these scenarios. In particular, only a small fraction 
(11 out of 193) of
the models found in~\cite{Lebedev:2008un} allow for supersymmetric vacua satisfying all the following constraints:\\
\phantom{.}\qquad{\it i)} both the hypercharge and an extra $\U1_X$ remain massless, \\
\phantom{.}\qquad{\it ii)} the extra $\U1_X$ is `hidden', i.e. all SM-particles have no $Q_X$ charge,\\
\phantom{.}\qquad{\it iii)} all exotic particles are decoupled at a scale $M_d$, close to the compactification scale,\\
\phantom{.}\qquad{\it iv)} at scales lower than $M_d$ there exist(s) some massless SM-singlet(s) with $Q_X\neq0$, 
         which can trigger the spontaneous breakdown of $\U1_X$ at an intermediate scale.\\
The subset of models leading to these properties share an additional feature: the unbroken gauge group after 
the action of the shift $V$ contains an \E6 factor, which is then broken down to the the SM gauge group times
an extra gauge sector by the Wilson lines. This means that these scenarios are favoured in models
with \E6 local GUTs, in the jargon of~\cite{Lebedev:2008un}.

\subsubsection{A promising string realisation of kinetic mixing}\label{SEC:Promising}

Let us now inspect the details of one example.  Consider the model defined by the shift vector
\beq
\label{eq:Vmodel2WL}
V~=~\tfrac1{6}(-2,-3,1,0,0,0,0,0)(0,0,0,0,0,0,0,0)
\eeq
and the Wilson lines
\begin{subequations}
\begin{eqnarray}
\mathcal{A}_2 &=&
\tfrac14(0,2,6,-10,-2,0,0,0)(5,-1,-5,-5,-5,-5,-5,5)\;,\\ 
\mathcal{A}_3 & = &
\tfrac16(-1,3,7,-5,1,1,1,1)(5,1,-5,-5,-5,-3,-3,3)\;,
\end{eqnarray}
\end{subequations}
satisfying eqs.~\eqref{eq:ModInv}. By itself, the shift $V$ is known as the \Z6--II 
standard (gauge) embedding and leads to the breaking 
$\E8\x\E8\rightarrow\E6\x\U1^2\x\E8$. After including both Wilson lines, the unbroken
gauge group is $\SU3_C\x\SU2_L\x\U1_Y\x[\SU8\x\U1_X\x\U1_\mathrm{anom}\x\U1^3]$. The 
4D $\maN=1$ matter spectrum is shown in Table~\ref{tab:spectrum}. The modular
group after compactification is 
$SL(2,\Z{})\x\Gamma_1(3)_{T_2}\x\Gamma_1(2)_{T_3}\x\Gamma^1(2)_{U_3}$.

The only relevant (as we shall shortly see) \U1 generators are given in the Cartan basis of \E8\x\E8 by
\begin{subequations}\label{eq:tsmodel2WL}
\begin{eqnarray}
t_Y &=& (0,0,0,-1/2,-1/2,1/3,1/3,1/3)(0,0,0,0,0,0,0,0)\,,\\
t_X &=& \tfrac1{4\sqrt2}(0,0,0,0,0,0,0,0)(1,-1,-1,-1,-1,3,3,3)\,,
\end{eqnarray}
\end{subequations}
where we have taken the phenomenologically favoured normalisation for 
the hypercharge $|t_Y|^2=5/6$. Since $\tr(Q_Y Q_X)=4\sqrt2\neq0$, 
there is a non-vanishing one-loop string contribution to the mixing between $\U1_Y$ and $\U1_X$
(see appendix~\ref{app:details} for further details):
\beq
\label{eq:Deltamodel2WL}
\Delta_{YX}= \frac1{16\pi^2} \frac{8\sqrt2}{3} \log \bigg(|\eta(3\,T_2)|^4 \mathrm{Im}(T_2)\bigg)\,,
\eeq
with $b_{YX}^2 =8\sqrt2$ and $b_{YX}^3 =0$. In this case, $\Delta_{YX}$
is about $1/40$ assuming that the modulus can be stabilised at $\langle T_2\rangle\sim1$.

\begin{table}[!t!]
\begin{center}
{\scriptsize
\begin{tabular}{|rlc|rlc|c|rlc|}
\cline{1-6}\cline{8-10}
  \#  &  Irrep                                          & Label & \# & Anti-irrep       & Label    &&  \#  & Irrep & Label \\
\cline{1-6}\cline{8-10}
  4 & $( {\bf 3}, {\bf 2};  {\bf 1})_{(1/6,\,0)}\phantom{A^{A^A	}}$   & $q_i$ &
  1 & $( {\bf\overline{3}}, {\bf 2};  {\bf 1})_{(-1/6,\,0)}$  & $\bar q_i$ &&
 45 & $( {\bf 1}, {\bf 1};  {\bf 1})_{(0,\,0)}$     & $s^0_i$ \\
 13 & $( {\bf 1}, {\bf 2};  {\bf 1})_{(-1/2,\,0)}$  & $\ell_i$ &
 10 & $( {\bf 1}, {\bf 2};  {\bf 1})_{(1/2,\,0)}$      & $\bar\ell_i$ && 
  4 & $( {\bf 1}, {\bf 1};  {\bf 1})_{(0,\,2\sqrt2/3)}$        & $\xi^+_i$ \\ 
  5 & $( {\bf\overline{3}}, {\bf 1}; {\bf 1})_{(-2/3,\,0)}$  & $\bar u_i$ &
  2 & $( {\bf 3}, {\bf 1};  {\bf 1})_{(2/3,\,0)}$      & $u_i$ && 
  4 & $( {\bf 1}, {\bf 1};  {\bf 1})_{(0,\,-2\sqrt2/3)}$        & $\xi^-_i$ \\ 
  5 & $( {\bf 1}, {\bf 1};  {\bf 1})_{(1,\,0)}$       & $\bar e_i$ &
  2 & $( {\bf 1}, {\bf 1};  {\bf 1})_{(-1,\,0)}$      & $e_i$  &&
  7 & $( {\bf 1}, {\bf 1};  {\bf\overline{8}})_{(0,\,-1/6\sqrt2)}$    & $\bar h_i$ \\
 10 & $( {\bf \overline{3}}, {\bf 1};  {\bf 1})_{(1/3,\,0)}$ & $\bar d_i$ &
  7 & $( {\bf 3}, {\bf 1};  {\bf 1})_{(-1/3,\,0)}$     & $d_i$ &&
  7 & $( {\bf 1}, {\bf 1};  {\bf 8})_{(0,\,1/6\sqrt2)}$        & $h_i$ \\ 
\cline{1-6}\cline{8-10}
  4 & $( {\bf 1},  {\bf 1};  {\bf 1})_{(1/2,\,\sqrt2/3)}$   & $s^{++}_i$ &  
  4 & $( {\bf 1},  {\bf 1};  {\bf 1})_{(-1/2,\,-\sqrt2/3)}$  & $s^{--}_i$ &  \multicolumn{3}{c}{$\phantom{I^{I^I}}$} \\
  4 & $( {\bf 1},  {\bf 1};  {\bf 1})_{(-1/2,\,\sqrt2/3)}$   & $s^{-+}_i$ &  
  4 & $( {\bf 1},  {\bf 1};  {\bf 1})_{(1/2,\,-\sqrt2/3)}$  & $s^{+-}_i$ &  \multicolumn{3}{c}{$\phantom{I^{I^I}}$} \\
  2 & $( {\bf 1},  {\bf 1};  {\bf\overline8})_{(1/2,\,1/2\sqrt2)}$   & $\sigma^{+}_i$ &  
  2 & $( {\bf 1},  {\bf 1};  {\bf 8})_{(-1/2,\,-1/2\sqrt2)}$  & $\bar\sigma^{-}_i$ &  \multicolumn{3}{c}{$\phantom{I^{I^I}}$} \\
  2 & $( {\bf 3},  {\bf 1};  {\bf 1})_{(1/6,\,2/\sqrt2)}$   & $w^+_i$ &  
  2 & $( {\bf\overline{3}}, {\bf 1};  {\bf 1})_{(-1/6,\,-2/\sqrt2)}$  & $\bar w^-_i$ & \multicolumn{3}{c}{$\phantom{I^{I^I}}$} \\
  2 & $( {\bf 3},  {\bf 1};  {\bf 1})_{(1/6,\,-2/\sqrt2)}$   & $w^-_i$ &  
  2 & $( {\bf\overline{3}}, {\bf 1};  {\bf 1})_{(-1/6,\,2/\sqrt2)}$  & $\bar w^+_i$ & \multicolumn{3}{c}{$\phantom{I^{I^I}}$} \\
  4 & $( {\bf 1},  {\bf 2};  {\bf 1})_{(0,\,\sqrt2/3)}$   & $m^+_i$ &
  4 & $( {\bf 1}, {\bf 2};   {\bf 1})_{(0,\,-\sqrt2/3)}$ &  $m^-_i$ & \multicolumn{3}{c}{$\phantom{I^{I^I}}$} \\
\cline{1-6}
\end{tabular}
\caption{Massless spectrum. Representations  with respect to
$[\SU3_C\times\SU2_L]\times[\SU8]$ are given  in  bold face, the hypercharge and the $\U1_X$ charge are
indicated as subscript.\label{tab:spectrum}}
}
\end{center}
\end{table}

To study the phenomenology of this model,
we choose the specific vacuum, in which only the fields
\begin{equation}
\label{eq:stilde}
 \{\widetilde{s}_i\}
 = \{
s^0_{1}, s^0_{2}, s^0_{3}, s^0_{12}, s^0_{19}, s^0_{22}, s^0_{29}, s^0_{32}, s^0_{36},
s^0_{38}, s^0_{46}, s^0_{48}, s^0_{50}, s^0_{51}, s^0_{57}
 \}  
\end{equation}
develop non-zero VEVs, while the expectation values of all other fields vanish.
The existence of the holomorphic monomial
\begin{equation}
\label{eq:monom}
\psi=
s^0_{1} s^0_{3} s^0_{36} s^0_{38}
(s^0_{2} s^0_{12} s^0_{19} s^0_{22} s^0_{29} s^0_{46} s^0_{48} s^0_{57})^2 
(s^0_{32} s^0_{50} s^0_{51})^4 
\end{equation}
ensures the cancellation of the Fayet--Iliopoulos term~\eqref{eq:FIterm}
and, thus, $\maN=1$ supersymmetry below the compactification scale.

Our choice of the vacuum~\eqref{eq:stilde} has further consequences.
First, all vector--like exotics attain large masses and decouple at 
$M_d\sim 0.1$ in string units. At the
same scale, all SM-singlets $s^0_i$, and all $h_i, \bar h_i, \xi^\pm_i$ but 
one pair of $(h_i,\bar h_j)$ and $(\xi^+_i,\xi^-_j)$ acquire masses.
Secondly, the gauge group is spontaneously broken down to
\begin{equation}
G_\mathrm{SM}\x[\SU8\x\U1_X]_\mathrm{hidden}\;,
\end{equation} 
where $\SU8\x\U1_X$ is hidden in the sense that no SM-field is
charged under this group. Note that the only surviving Abelian symmetries
correspond to the generators given in~\eqref{eq:tsmodel2WL}.
Therefore, the vacuum chosen contains only the spectrum
of the MSSM plus the two pairs of multiplets $(h_i,\bar h_j)$ and $(\xi^+_i,\xi^-_j)$,
both of which are charged under $\U1_X$. Let us call these SM singlets $(h_+,h_-)$
and $(\xi_+,\xi_-)$ respectively.

It follows that $b_{YX}=0$ below $M_d$ and, therefore,~\eqref{eq:chirun} becomes
\beq
\frac{\chi_{\mbox{\tiny $YX$}}}{g_Y g_X} = \frac{4\sqrt2}{16\pi^2} \log \frac{M^2_S}{M_d^2} + \Delta_{YX}
\eeq
and does not run. 
Consequently this has nearly the correct ingredients for an interesting hidden sector kinetically
mixing with the hypercharge: the kinetic mixing is present and we have some hidden vector-like matter.
Such matter can cause higgsing of the hidden gauge group and may be interesting for dark-matter 
phenomenology or laboratory experiments at the low energy, high intensity frontier, as emphasized in the introduction.

\section{Phenomenology of hidden photons from heterotic orbifolds}
\label{sec:phenohiddenphotons}

Here we discuss the different \emph{hidden} \U1 phenomenology and 
the predictions from (or implications for a discovery on)
heterotic orbifolds.

The low energy limits of heterotic orbifolds provide consistent, 
complete and calculable realisations of supersymmetric field theories. 
They therefore provide meaningful restrictions on the phenomenology that
 we can obtain. Specifically with regard to supplementary \U1 symmetries, 
in addition to the limit of the maximum total rank of all gauge groups, 
there are many further constraints. The most important features for phenomenology 
are whether there is hidden matter (such as in the example above) and, if so, 
what its couplings are; and 
whether the gauge boson has a mass. To obtain massive non-anomalous \U1s in 
heterotic orbifolds we require a spontaneous breaking mechanism. This is
 because the theory, in contrast to D-brane models, lacks the axions to give 
them St\"uckelberg masses. Moreover, only one anomalous \U1 is allowed, so 
there is no possibility of using a fermion condensate to give masses to any others. 

In what follows, we shall assume that the hidden sectors play no role 
in supersymmetry breaking, and we shall assume that moduli stabilisation 
and the integrating out of massive singlets has taken place. Thus we shall 
treat the resulting theory as a softly broken globally supersymmetric model, possibly with 
supersymmetry-breaking masses of similar order to those in the visible sector 
(in the standard case of gravity mediation) or much smaller masses (for gauge-mediated scenarios, 
see for example \cite{Brummer:2011yd}).

\subsection{Massless hidden \U1s}

A massless hidden \U1 can be interesting phenomenologically in a supersymmetric 
theory due to its gaugino. The key issue in this case is that in heterotic 
orbifolds the order of magnitude of the hidden gauge coupling and the kinetic
mixing (if present) are fixed, to being within roughly an order of magnitude or so of the standard model couplings 
and $10^{-3}$ respectively. As discussed in \cite{Ibarra:2008kn,Goodsell:2010ie}, 
if the hidden gaugino is the LSP then it will be overproduced, as the mixing cannot 
be reduced sufficiently to avoid this. This fate can be avoided in, for example, 
models with gauge mediation \cite{Brummer:2011yd} where the hidden gaugino can 
decay to a gravitino. In that case, there could be signals due to displaced 
vertices at the LHC \cite{Arvanitaki:2009hb}. 

Alternatively, in (the much more standard case of) models with gravity
 mediation, the difficulty could potentially be avoided by allowing for 
hidden matter. However, this would then possess millicharges under the 
hypercharge, and for mixing of order $10^{-3}$ is constrained to have masses above about $100$ MeV \cite{Davidson:2000hf}.
Since the 
theory is supersymmetric, there would necessarily be hidden fermions, and 
so there would need to be a hidden supersymmetric mass for these (either 
an explicit hidden $\mu$-term or a form of hidden Higgs mechanism that 
does not break the \U1). Once this is allowed for, however, we could hope 
to detect the hidden gauginos in collider experiments as above. However, 
the scenario would not be interesting for dark matter experiments since we 
cannot substantially reduce the hidden gauge coupling; the self-interactions 
of a particle charged under the hidden \U1 would be too strong, violating observations 
about the clustering of dark matter \cite{Ackerman:2008gi} (and a particle 
not charged under the hidden \U1 would interact too weakly with the visible 
sector to be detected). So then we should simply ensure that the relic density from the hidden sector is small.

A hidden Dirac fermion $\psi$ with mass $m_{D}$ and hidden gauge coupling 
$g_h$ is thermalised provided that the rate of production is greater 
than the Hubble constant at some point. Assuming that the temperature is 
at some time above the mass of the hidden fermion, we obtain (roughly) 
\begin{align}
1 < \frac{\Gamma}{H} \sim (g_Y g_h \chi)^2 \left(\frac{M_P}{m_D}\right)\,,
\end{align}
which implies that most such hidden sectors at experimentally accessible energies in heterotic orbifolds are thermalised, since the gauge coupling and kinetic mixing (if present and supersymmetric) cannot be substantially reduced in magnitude.\footnote{Note that, even if the above bound is not met the hidden sector could become thermalised through decays of moduli etc, although this is somewhat model dependent. }

For such a thermal species, the relic density is approximately given by 
\begin{align}
\frac{\Omega_\psi h^2}{0.112} \approx 10^{-4} \left( \frac{0.1}{g_h} \right)^4 \left( \frac{m_{D}}{\mathrm{GeV}}\right)^2 \ll 1,
\end{align}
implying that the hidden matter cannot be too heavy. Including the constraints 
on millicharges and allowing for a hidden gauge coupling as large as $1$, we then 
constrain the hidden matter to roughly lie in the range 
\begin{equation}
100\  \mathrm{MeV} < m_D \lesssim 10^4\ \mathrm{GeV}\,,
\end{equation}
although the upper bound could be avoided if the reheating temperature is low.

\subsection{Hidden \U1 masses through supersymmetric breaking}

Heterotic orbifold models typically begin with several \U1 factors that are broken 
supersymmetrically by the VEVs of standard-model singlets. These VEVs are induced 
by the effective Fayet-Iliopoulos term corresponding to the one anomalous \U1, and 
their exact values depend on the details of moduli stabilisation. These expectation
values are expected to be of the order of $0.1 M_S$, and the \U1 groups directly 
broken in this way will thus have very large masses - so the number surviving at 
low energies is typically small. However, we could in principle obtain supersymmetric 
breaking of a hidden \U1 with a small mass: a prototypical example of such breaking 
would be a hidden-sector theory with three fields $S_h, H_+$ and $H_-$ which have 
charges $0, 1$ and $-1$ under the hidden \U1 respectively, and superpotential
\begin{align}
W \supset \lambda S_h ( H_+ H_- - \mu^2).
\end{align}
This theory spontaneously generates a vacuum expectation value for $H_\pm$, giving a 
hidden photon mass $2 g_h \mu$, and together with the $D$-term potential gives masses 
to all of the scalars and fermions. If $\lambda \sim g_h$ then these are all of order 
the hidden photon mass. Of course, we do not expect to obtain a tadpole term directly 
in the orbifold: $\mu$ should be regarded as effective, either arising from a term of 
the form $M_S^2 s_0^n S_h$, where $s_0$ is some field that obtains a string-scale VEV 
and $n$ is suitably high ($\sim 32$ for $10$ GeV hidden gauge bosons!) or more realistically 
arising from the effect of strong gauge dynamics; since the hidden sector of heterotic 
orbifolds typically includes a non-Abelian group, this can be used to induce supersymmetric 
breaking as above if there is light hidden matter. This does, however, potentially preclude 
the use of the non-Abelian group for moduli stabilisation - in order to effect gauge 
symmetry breaking we assume that the moduli are stabilised already (there could be more 
than one non-Abelian hidden group, for example, with only one of them charged under some 
hidden light matter). 

\subsubsection{An example}
\label{SEC:toyexample}

With the above assumption, an example of strong gauge dynamics breaking a hidden \U1 
which could appear from a heterotic orbifold would be to have a hidden sector consisting 
of four fields $\psi, \tilde{\psi}, \phi, \tilde{\phi}$ which are charged under \SU{N}\x\U1. 
$\psi, \phi$ are fundamentals of \SU{N} and have charges $0, +1$ respectively under the \U1; 
$\tilde{\psi}, \tilde{\phi}$ are antifundamentals under the \SU{N} and have charges $0, -1$ 
under the \U1. We take the perturbative effective superpotential
\begin{align}
W_{\rm pert} =~& \frac{\lambda}{M_S} (\phi \tilde{\psi}) (\psi \tilde{\phi})\,,
\end{align}
where we have used brackets to show how the \SU{N} indices should be contracted. 
Below the strong coupling scale of the \SU{N}, the fields condense into  the 
(matrix of) mesons $M\equiv (\phi,\psi) \otimes (\tilde{\phi},\tilde{\psi})^T$. 
For \SU{N} with $N_f$ flavours the meson transforms under the adjoint of \SU{N_f} 
and we have the classic ADS 1PI superpotential (see e.g. \cite{Intriligator:1995au} for a good review):
\begin{align}
W =~& (N - N_f) \left( \frac{\Lambda^{3N - N_f}}{\mathrm{det} M} \right)^{\frac{1}{N-N_f}} + W_{\rm pert}\,,
\end{align}
where $ W_{\rm pert}$ is the perturbative potential, given in terms of the gauge invariants $M$. 
In the present example, we conveniently define 
$U \equiv  (\phi \tilde{\phi}), V \equiv (\psi \tilde{\psi}), 
 H_+ \equiv (\phi \tilde{\psi}), H_- \equiv (\psi \tilde{\phi})$ 
so that $\mathrm{det} M = UV-H_+ H_-$, arriving at
\begin{align}
W =~& (N - 2) \left( \frac{\Lambda^{3N - 2}}{UV-H_+ H_-} \right)^{\frac{1}{N-2}} + \frac{\lambda}{M_S} H_+ H_- .
\end{align}
The D-term equations enforce $H_+ =e^{i\theta} H_-$ for some phase $\theta$; choosing 
the VEV of $H_+$ to be real fixes $\theta$ and we solve the F-term equations for the 
above to give (writing $M_S \equiv z\, \Lambda,\, z \gg 1 $):
\begin{align}
\bra H_+ \ket =~& \Lambda^2 \left( \frac{z}{\lambda} \right)^{\frac{N-2}{2(N-1)}}.
\end{align} 
Since now $H_+$ has dimension two, the canonical field is found by dividing by some scale of order 
of the condensation scale. This produces the hidden-photon mass
\begin{align}
m_{\gamma'} \sim~& 2 g_h \Lambda \left( \frac{z}{\lambda} \right)^{\frac{N-2}{2(N-1)}} \nn\\
\sim~& 2 g_h M_S\, z^{- \frac{N}{2(N-1)}} \lambda^{-\frac{N-2}{2(N-1)}} .
\end{align}
Here $ g_h$ is the coupling of the hidden \U1, and we shall write $g_N$ for the \SU{N} 
gauge coupling with $\alpha_N \equiv g_N^2/(4\pi)$. The exponent of $z$ will thus vary 
between $-1$ and $-1/2$, so the scale of breaking is set by the condensation scale, given by (using $b \equiv 3N - N_f$)
\begin{align}
\Lambda =~& M_D \exp\left(- \frac{8\pi^2}{b\, g^2_N (M_D)}\right) \nn\\
z=~&\frac{M_S}{M_D}\exp\left(\frac{2\pi}{b\, \alpha_N (M_D)}\right)\,,
\end{align}
where $M_D$ is the scale of the lightest heavy particle charged under the \SU{N} 
that is integrated out (as mentioned previously, these are generically present 
in heterotic orbifolds). Interestingly, however, the hidden \U1 coupling will become 
weaker due to the extra matter; we have
\begin{align}
\frac{1}{g_h^2 (\Lambda)} =~& \frac{1}{g_h^2 (M_D)} - \frac{2N}{8\pi^2} \log \Lambda/M_D.
\end{align}
For some sample values, let us suppose that there is a large amount of matter 
above $M_D = 10^{16}$ GeV so that $g_N (M_D)$ is small. Taking $N=3$ and $\alpha_N^{-1} (M_D) =56$ 
we find a hidden gauge boson mass of $1$ GeV~\footnote{We also take $\lambda \sim 1$ and $g_h \sim 0.3$.}. 
Of course, such a 
value for the non-Abelian coupling is rather weak; however, this problem could be avoided by, 
for example, introducing an extra flavour that has an intermediate scale mass. In this way, 
the coupling would ``walk'' down from $M_D$ to this new scale, before becoming stronger - 
reducing the condensation scale. Alternatively, it would be interesting to consider an 
ISS-like model (by adding more light flavours) which would potentially also break supersymmetry. 
However, the model is modified, though, by invoking strong dynamics to break the Abelian gauge 
symmetry. We will find that the gauge boson mass depends exponentially on the 
scales and couplings, allowing a wide range of values and phenomenology.

\subsection{Dark Forces after supersymmetry breaking}

Most Dark Force models constructed in the literature break the hidden \U1 
in the low energy theory after supersymmetry is broken. This can be induced 
by the electroweak symmetry breaking in the visible sector, where the Higgs 
expectation values give a D-term to the hypercharge, and this is communicated 
to the hidden sector via the kinetic mixing \cite{Morrissey:2009ur}. In the 
context of gauge mediation there are  many possibilities for models, since 
the soft masses in the hidden sector can be naturally small, but in the context 
of gravity mediation this would require the hidden sector to be sequestered. 
Hence for heterotic orbifolds, as investigated in \cite{Andreas:2011in}, an 
ideal scenario would involve \emph{radiative} breaking of the hidden gauge 
group, as could be achieved in a simple hidden sector with three fields 
$S, H_+, H_-$ and superpotential\footnote{This was first proposed as a Dark 
Force model in the context of gauge mediation in \cite{Hooper:2008im} and 
explored in more detail in \cite{Morrissey:2009ur}.}
\begin{align}
W \supset ~& \lambda_S S\, H_+ H_-.
\end{align} 
This has the advantage of having no scales, and that we actually require 
$\lambda_S$ to be not suppressed. Unfortunately we were so far not able to 
find examples in this class, and we leave this as a challenge for future work. 

The simplest supersymmetric Dark Force model would have superpotential
\begin{align}
W \supset ~& \mu\, (\xi_+ \xi_-), 
\end{align}
where we require $\mu \lesssim 10$ GeV and there to be a hidden gaugino 
mass (ideally of similar magnitude). The kinetic mixing and hypercharge 
D-term would then break the hidden gauge group spontaneously \cite{Morrissey:2009ur}.

\subsubsection{A promising example}

The model of section \ref{SEC:Promising} can in principle exactly 
realise a Dark Force scenario. This is because it contains a perturbative superpotential with terms 
\begin{align}
W \supset~& s_0^5 \frac{1}{M_S}(\xi_+ \xi_-) (h_+ h_-)  + s_0^8\, M_S\, (h_+ h_-)\,,
\end{align}
where $s_0\sim\mathcal{O}(0.1)$ is the VEV of a $G_\mathrm{SM}\x\U1_X$ singlet expressed in string units, and 
$h_\pm$ are SM singlets charged under both $\U1_X$ and a hidden non-Abelian 
group. There are also additional heavy fields 
$h_i, \bar{h}_i, \xi^+_i, \xi^-_i$ which will slow the running of the non-Abelian group above
$M_D$, and (if some of the $h_i, \bar{h}_i$ obtain VEVs) can break it down to \SU5. So we shall 
analyse an \SU{N} sector with superpotential
\begin{align}
W \supset~& s_0^n \frac{1}{M_S}(\xi_+ \xi_-) (h_+ h_-)  + s_0^m M_S (h_+ h_-).
\end{align}
As in subsection \ref{SEC:toyexample}, at the scale $\Lambda$ there is a dynamically generated superpotential given in terms of the gauge invariant $M\equiv h_+ h_-$:
\begin{align}
W \supset (N-1)\left( \frac{\Lambda^{3N-1}}{M} \right)^{1/(N-1)} + s_0^m M_S  M + s_0^n \frac{1}{M_S}(\xi_+ \xi_-) M.
\end{align}
$M$ then obtains a VEV (writing $M_S= z \Lambda$, as before) of size
\begin{align}
M= \Lambda^2 (s_0^m z)^{-\frac{N-1}{N}}.
\end{align}
We find thus an effective superpotential of the form $W = \mu (\xi_+ \xi_-)$, with
\begin{align}
\mu =~& s_0^n \frac{\Lambda^2}{M_S} (s_0^m z)^{-\frac{N-1}{N}} \nn\\
=~& M_S s_0^{n - m(\frac{N-1}{N})} z^{-\frac{3N-1}{N}} . 
\end{align}
Taking the model of section  \ref{SEC:Promising} (i.e. $m=8 ,n=5$) we find that, in order to obtain 
a $\mu$ of $10$ GeV, we need $z\sim 10^7$, corresponding to a condensation scale of $10^{11} $ GeV. 
If, as mentioned above, we break the gauge group down to \SU5 at a scale $M_D \sim 10^{16}$ GeV then 
we can take $\alpha_N^{-1} (M_D) = 24$ to  realise a Dark Force model.

\section{Discussion}
\label{sec:discussion}

Hidden forces might play an important
role in nature, most notably in the context of dark matter. To probe the existence of Abelian hidden
forces, the magnitude of their kinetic mixing with the hypercharge in the proposed models must be 
contrasted to currently observed bounds. On the other hand, string theory offers a remarkable playground 
for these new interactions, since additional \U1s generically appear in all kinds of string compactifications.
Unfortunately, kinetic mixing has been studied only in some particular scenarios and in some cases only
vanishing mixing has been found, which renders the new forces undetectable, thus irrelevant for nature.

In this paper, we aimed at improving this situation. We have studied the kinetic mixing between \U1s
in a special class of string constructions: heterotic orbifold models. We have noticed that the
computation of kinetic mixing contribution $\Delta_{ab}$ here is akin to the computation of threshold corrections.
In particular, the result mostly depends on the massless modes of the $\maN=2$ subsectors of the orbifold.
This constrains the candidates that exhibit mixing to \Z{N} with non-prime $N$ and \Z{N}\x\Z{M} 
orbifold models. Using the resulting expression for kinetic mixing 
we have been able to provide a simple formula that helps one to recognize promising candidates
with non-vanishing mixing. We also found that the resulting size of kinetic mixing depends on the size of the compact
space, which we assumed to have been stabilised.

As an application of our previous results, we have explored explicit \Z6--II orbifold models with 
and without Wilson lines. In the former case, despite 
not having realistic gauge symmetries, we find 10 out of a total of 61 orbifold models with 
Abelian kinetic mixing in the interval $10^{-2} \lesssim \Delta_{ab} \lesssim 10^{-1}$, values which are consistent
with previous expectations~\cite{Dienes:1996zr}. It is interesting to note that, although this result resembles
the one obtained previously in the context of type II strings with D3-branes in orbifold backgrounds~\cite{Bullimore:2010aj}, 
the models studied here are globally consistent constructions. This can also be contrasted to the toy example 
with intersecting D6-branes in~\cite{Abel:2008ai}. 

The \Z6--II orbifold models with (2 and 3)  Wilson lines that we have investigated are those of the Mini-Landscape,
which possess many properties of the (N)MSSM. We found that 255 (out of 267) models have non-trivial mixing
between the hypercharge and at least one additional $\U1_X$. The size of the stringy contribution to the 
kinetic mixing lies in the range $10^{-4}\,-\,10^{-2}$. This is in contrast to previous results~\cite{Dienes:1996zr}, 
where semi-realistic orbifold models had been studied and no model with kinetic mixing was found.
However, demanding the new Abelian symmetry to be truly hidden and to mix with the hypercharge in 
supersymmetric vacua turned out to be more challenging: only 11 models survive these new demands.

We have also worked out explicitly the details of a sample model. We provided a model with a supersymmetric vacuum, 
observable gauge group $\SU3_C\x\SU2_L\x\U1_Y$ and hidden $\U1_X$, and the matter spectrum of the MSSM plus
a couple of vectorlike SM singlets that mediate interactions between both sectors. The mixing of the 
hypercharge and the hidden \U1 occurs only in one $\maN=2$ sector, rendering the result $\Delta_{YX}\sim\frac1{40}$ stable
against the running of the couplings.

Finally, we have discussed some phenomenologically appealing possibilities that our models allow. In particular, with 
the model that we have used as an example, we showed that a Dark Force scenario arises due to the existence of
a strongly interacting sector and adequate hidden matter couplings. This situation is generic, even though the 
precise details of the form of the couplings and the condensation scale depend on the particulars of each model.

Our study can be extended in several ways. First, we have focused on \Z6--II orbifolds and ignored other promising
models, such as \Z{12}--II and \Z2\x\Z2, which have shown to be (at least) as promising as \Z6--II~\cite{Kim:2006hv,Blaszczyk:2009in}. 
In a fast test, we could verify that there are plenty of models similar to those of the Mini-Landscape with kinetic mixing. 
It would be worth studying the details; in particular the phenomenology of the hidden sectors and hidden matter could 
be very interesting. Secondly, in this work we have concentrated on truly hidden \U1s, although
$Z'$ symmetries appear more frequently. A careful study of the kinetic mixing and other properties of these symmetries
will be carried out elsewhere. It would also be interesting to perform our analysis avoiding the assumption of
moduli stabilisation, i.e. in heterotic orbifold models with stabilised moduli.

\section*{Acknowledgements}
We thank O.~Lebedev and J.~Erler for helpful discussions. M.G. was supported by SFB grant 676
and ERC advanced grant 226371.
S.~R.-S. was partially supported by CONACyT project 82291 and DGAPA project IA101811.

\appendix
\section{Further details on a promising model}
\label{app:details}

We provide here the data of the model presented in section~\ref{SEC:Promising}, which allows one
to compute the magnitude of kinetic mixing.

\paragraph{\bs{\Z2} $\bs{\maN=2}$ theory.}
The unbroken gauge group is $\SU4\x\SU3\x\U1^3\x[\SU8\x\U1]$, where we have used squared parenthesis 
in the subgroups arising from the second \E8 of the original heterotic string. The $\maN=2$ \U1 generators
are given by
\begin{eqnarray}
\label{eq:Z2Neq2U1s}
t_{13'}&=&\tfrac{1}{\sqrt{30}}(-1,5,-1,0,0,1,1,1)(0,0,0,0,0,0,0,0)\,,\\
t_{14'}&=&\tfrac{1}{\sqrt{20}}(-1,0,4,0,0,1,1,1)(0,0,0,0,0,0,0,0)\,,\nonumber\\
t_{15'}&=&\tfrac{1}{\sqrt{2}}(0,0,0,1,1,0,0,0)(0,0,0,0,0,0,0,0)\,,\nonumber\\
t_{16'}&=&\tfrac{1}{\sqrt{32}}(0,0,0,0,0,0,0,0)(1,-1,-1,-1,-1,3,3,-3)\,.\nonumber
\end{eqnarray}

Listing complete hypermultiplets, the $\maN=2$ matter spectrum contains the following gauge representations\\
\centerline{\small
\begin{tabular}{|ccc||ccc|}
\hline
\# & Irrep & \U1 charges & \# & Irrep & \U1 charges\\
\hline
2 & ( \bs1,\,\bs3,\,\bs1) & {\scriptsize$(-\frac{1}{\sqrt{30}},\,\frac2{\sqrt5},\,-\frac1{\sqrt2},\,0)$}&
8 & ( \bs1,\,\bs3,\,\bs1) & {\scriptsize$(\sqrt{\frac2{15}},\,\frac1{\sqrt5},\,\frac1{\sqrt2},\,0)$}\\ 
2 & ( \bs6,\,\bs1,\,\bs1) & {\scriptsize$(-\sqrt{\frac3{10}},\,\frac1{\sqrt5},\,\frac1{\sqrt2},\,0)$}&
8 & ( \bs1,\,\bs1,\,\bs8) & {\scriptsize$(0,\,0,\,\frac1{\sqrt2},\,-\frac1{\sqrt8})$}\\
8 & ( \bs4,\,\bs1,\,\bs1) & {\scriptsize$(\sqrt{\frac3{10}},\,\frac3{\sqrt{20}},\,0,\,0)$}&&&\\ 
\hline
\end{tabular}}
\vskip1mm
Consequently, the $\beta$ function coefficients are given by
$b_{\SU4}^{\maN=2}=b_{\SU3}^{\maN=2}=4$,
$b_{\SU8}^{\maN=2}=-8$ and
\begin{equation}
b_{\U1}^{\maN=2} =~\mbox{\scriptsize$
\left(
\begin{array}{cccc}
\frac{83}{10}&\frac{12}5\sqrt6&\frac32\sqrt{\frac35}&0\\
\frac{12}5\sqrt6&\frac{66}5&6\sqrt{\frac25}&0\\
\frac32\sqrt{\frac35}&6\sqrt{\frac25}&\frac{53}2&-8\\
0&0&-8&4
\end{array}
\right)\,.
$}
\end{equation}
In the notation of eq.~\eqref{eq:Neq2combination}, the overlap of the hypercharge and $\U1_X$ is given by
\begin{subequations}
\label{eqs:PromisingOverlap}
\begin{eqnarray}
 &\mbox{\small
  $n_Y = (\tfrac1{\sqrt{30}},\,\tfrac1{\sqrt{20}},\,-\tfrac1{\sqrt{2}},\,0),\,
  m_Y^{\SU4} = (\tfrac1{12},\,\tfrac1{6},\,\tfrac1{4}),\, 
  m_Y^{\SU3} = (-\tfrac16,-\tfrac13),\,
  m_Y^{\SU8}=0$} &\\
 &n_X~=~(0,\,0,\,0,\,1)\,,\quad\qquad m_X^{b'}=0 \text{ for all }b'&
\end{eqnarray}
\end{subequations}
Thus, according to eq.~\eqref{eq:mixingsimple}, $b_{YX}^2=8\sqrt2$.

\paragraph{\bs{\Z3} $\bs{\maN=2}$ theory.}
The unbroken gauge group is $\SU4\x\SU2_a\x\SU2_b\x\U1^3\x[\SU8\x\U1]$, where, as before, the parethesis refer
to the subgroups arising from the second \E8 of the original heterotic string. The $\maN=2$ \U1 generators are
\begin{eqnarray}
\label{eq:Z3Neq2U1s}
t_{13'}&=&\tfrac{1}{4\sqrt{11}}(11,-5,5,1,1,1,1,1)(0,0,0,0,0,0,0,0)\,,\\
t_{14'}&=&\tfrac{1}{\sqrt{66}}(0,6,5,1,1,1,1,1)(0,0,0,0,0,0,0,0)\,,\nonumber\\
t_{15'}&=&\tfrac{1}{\sqrt{32}}(0,0,0,0,0,0,0,0)(1,-1,-1,-1,-1,3,3,-3)\,.\nonumber\\
t_{16'}&=&\tfrac{1}{\sqrt{6}}(0,0,-1,1,1,1,1,1)(0,0,0,0,0,0,0,0)\,,\nonumber
\end{eqnarray}

Listing complete hypermultiplets, the $\maN=2$ matter spectrum contains the following gauge representations\\
\centerline{\small
\begin{tabular}{|ccc||ccc|}
\hline
\# & Irrep & \U1 charges & \# & Irrep & \U1 charges\\
\hline
1 & ( \bs6,\,\bs1,\,\bs1,\,\bs1) & {\scriptsize$(\frac3{\sqrt{11}},\,\frac1{\sqrt{66}},\,0,\,\frac1{\sqrt6})$}&
3 & ( \bs1,\,\bs1,\,\bs1,\,\bsb8) & {\scriptsize$(0,\,0,\,-\frac{1}{6 \sqrt{2}},\,\sqrt{\frac{2}{3}})$}\\
1 & ( \bs1,\,\bs2,\,\bs2,\,\bs1) & {\scriptsize$(\frac1{\sqrt{11}},\,\sqrt{\frac8{33}},\,0,\, \sqrt{\frac23})$}&
3 & ( \bs1,\,\bs1,\,\bs1,\,\bsb8) & {\scriptsize$(\frac{2}{\sqrt{11}},\,-\sqrt{\frac{3}{22}},\,-\frac{1}{6 \sqrt{2}},\,-\frac{1}{\sqrt{6}})$}\\
3 & ( \bs6,\,\bs1,\,\bs1,\,\bs1) & {\scriptsize$(\frac1{3\sqrt{11}},\,-\frac7{3\sqrt{66}},\,0,\,\frac5{3\sqrt6})$}&
6 & ( \bs1,\,\bs1,\,\bs1,\,\bs1) & {\scriptsize$(-\frac{8}{3 \sqrt{11}},\,-\frac53 \sqrt{\frac{2}{33}},\,0,\,\frac13\sqrt{\frac{2}{3}})$}\\
3 & ( \bs4,\,\bs2,\,\bs1,\,\bs1) & {\scriptsize$(-\frac7{6\sqrt{11}},\,\frac43\sqrt{\frac2{33}},\,0,\,\frac13\sqrt{\frac2{3}})$}&
3 & ( \bs1,\,\bs1,\,\bs1,\,\bs1) & {\scriptsize$(-\frac{8}{3 \sqrt{11}},\,\frac{23}{3 \sqrt{66}},\,0,\,-\frac{1}{3 \sqrt{6}})$}\\
3 & ( \bs4,\,\bs1,\,\bs2,\,\bs1) & {\scriptsize$(-\frac1{6\sqrt{11}},\,-\frac{13}{3\sqrt{66}},\,0,\,-\frac{1}{3 \sqrt{6}})$}&
3 & ( \bs1,\,\bs1,\,\bs1,\,\bs1) & {\scriptsize$(-\frac{2}{3 \sqrt{11}},\,-\frac{19}{3 \sqrt{66}},\,0,\,-\frac{7}{3 \sqrt{6}})$}\\
3 & ( \bs1,\,\bs2,\,\bs2,\,\bs1) & {\scriptsize$(-\frac{5}{3 \sqrt{11}},\,\frac13\sqrt{\frac2{33}},\,0,\,-\frac23\sqrt{\frac2{3}})$}&
3 & ( \bs1,\,\bs1,\,\bs1,\,\bs1) & {\scriptsize$(-\frac{4}{3 \sqrt{11}},\,-\frac{5}{3 \sqrt{66}},\,-\frac{2 \sqrt{2}}{3},\,-\frac{5}{3 \sqrt{6}})$}\\
3 & ( \bs1,\,\bs1,\,\bs1,\,\bs8) & {\scriptsize$(\frac{2}{3 \sqrt{11}},\,-\frac73\sqrt{\frac2{33}},\,\frac1{6\sqrt2},\,\frac23\sqrt{\frac2{3}})$}&
3 & ( \bs1,\,\bs1,\,\bs1,\,\bs1) & {\scriptsize$(-\frac{2}{\sqrt{11}},\,\sqrt{\frac{3}{22}},\,\frac{2 \sqrt{2}}{3},\,-\frac{1}{\sqrt{6}})$}\\
3 & ( \bs1,\,\bs1,\,\bs1,\,\bs8) & {\scriptsize$(\frac{2}{3 \sqrt{11}},\,\frac{19}{3 \sqrt{66}},\,\frac{1}{6 \sqrt{2}},\,\frac{1}{3 \sqrt{6}})$} &&&\\
\hline
\end{tabular}
}
\vskip1mm
Consequently, the $\beta$ function coefficients are given by $b_{\SU4}^{\maN=2}=12$,
$b_{\SU2_a}^{\maN=2}=b_{\SU2_b}^{\maN=2}=16$,
$b_{\SU8}^{\maN=2}=-4$ and
\begin{equation}
b_{\U1}^{\maN=2} =~\mbox{\scriptsize$
\left(
\begin{array}{cccc}
\frac{490}{33}&-\frac{131}{33}\sqrt{\frac23}&-\frac43\sqrt{\frac2{11}}&\frac{13}3\sqrt{\frac2{33}}\\
-\frac{131}{33}\sqrt{\frac23}&\frac{2171}{99}&\frac{28}{3\sqrt{33}}&-\frac{25}{9\sqrt{11}}\\
-\frac43\sqrt{\frac2{11}}&\frac{28}{3\sqrt{33}}&\frac{10}3&\frac4{3\sqrt3}\\
\frac{13}3\sqrt{\frac2{33}}&-\frac{25}{9\sqrt{11}}&\frac4{3\sqrt3}&\frac{227}9
\end{array}
\right)\,.
$}
\end{equation}
In the notation of eq.~\eqref{eq:Neq2combination}, the overlap of the hypercharge and $\U1_X$ is given by
\begin{subequations}
\label{eqs:PromisingOverlap}
\begin{eqnarray}
 &m_Y^{\SU4} = (\tfrac16,\,\tfrac13,\,\tfrac12),\quad m_Y^{\SU2_b}=-\tfrac12\,,&\qquad n_Y=0,\,m_Y^{b'}=0 \text{ for other }b'\,,\\
 &n_X~=~(0,\,0,\,1,\,0)\,,&\qquad m_X^{b'}=0 \text{ for all }b'\,.
\end{eqnarray}
\end{subequations}
Thus, according to eq.~\eqref{eq:mixingsimple}, $b_{YX}^3=0$.

\section{Other promising models with kinetic mixing}
\label{app:moremodels}
The models presented here exhibit kinetic mixing. However, they do not satisfy all the qualities we demand in 
section~\ref{sec:semirealistic}.

\subsection{Example 1. Mixing in the observable \E8}
Another interesting example of \U1 kinetic mixing in heterotic orbifolds arises from considering the
\Z6--II gauge embedding:
\begin{subequations}
\begin{eqnarray}
V &=&\tfrac1{6}(2,-3,-3,0,0,0,0,0)(3,-1,-3,-3,-3,-3,-3,3)\;,\\
\mathcal{A}_2 &=&
\tfrac12(1,0,0,0,-2,-1,-1,1)(1,-1,-1,-1,-1,-1,1,3)\;,\\ 
\mathcal{A}_3 & = &
\tfrac13(3,-6,-4,-4,-4,-4,-4,5)(6,-3,-5,-5,-5,-5,-8,1)\;.
\end{eqnarray}
\end{subequations}
The interesting \U1 generators are given in the Cartan basis of \E8\x\E8 by
\begin{subequations}\label{eq:tsExample1}
\begin{eqnarray}
t_Y &=& (0,0,0,-1/2,-1/2,1/3,1/3,1/3)(0,0,0,0,0,0,0,0)\,,\\
t_X &=& (0,1,0,0,0,0,0,0)(0,0,0,0,0,0,0,0)\,.
\end{eqnarray}
\end{subequations}
The matter spectrum can be obtained from~\cite{WebTables:2006xx} 
or by using the program {\small\tt{orbifolder}}~\cite{Nilles:2011aj,ORB11}.

Using the data below, the resulting kinetic mixing is
\beq
\Delta_{YX}= -\frac1{4\pi^2}\log \bigg(|\eta(2\,T_3)\eta(U_3/2)|^4 \mathrm{Im}(T_3)\mathrm{Im}(U_3)\bigg)\,.
\eeq

\paragraph{\bs{\Z2} $\bs{\maN=2}$ theory.}
Using the notation of the previous appendix, the unbroken gauge group is $\SU5\x\U1^4\x[\SU6\x\SU2_a\x\SU2_b\x\U1]$, 
and the $\maN=2$ \U1 generators are given by
\begin{eqnarray}
\label{eq:Z2Neq2U1sEx1}
t_{12'}&=&\tfrac{1}{\sqrt{70}}(7,3,3,0,0,-1,-1,-1)(0,0,0,0,0,0,0,0)\,,\\
t_{13'}&=&\tfrac{1}{2\sqrt{7}}(0,4,-3,0,0,1,1,1)(0,0,0,0,0,0,0,0)\,,\nn\\
t_{14'}&=&\tfrac{1}{\sqrt{2}}(0,0,0,1,1,0,0,0)(0,0,0,0,0,0,0,0)\,,\nonumber\\
t_{15'}&=&\tfrac{1}{2}(0,0,1,0,0,1,1,1)(0,0,0,0,0,0,0,0)\,,\nonumber\\
t_{16'}&=&\tfrac{1}{\sqrt{6}}(0,0,0,0,0,0,0,0)(0,0,1,1,1,1,1,1)\,.\nonumber
\end{eqnarray}

The $\maN=2$ matter spectrum contains the following hypermultiplet gauge representations\\
\centerline{\small
\begin{tabular}{|ccc||ccc|}
\hline
\# & Irrep & \U1 charges & \# & Irrep & \U1 charges\\
\hline
2 & ( \bs5,\,\bs1,\,\bs1,\,\bs1)    & {\scriptsize$(\tfrac3{\sqrt{70}},\,-\tfrac3{2\sqrt7},\,\tfrac1{\sqrt2},\,-\tfrac12,\,0)$}&
1 & ( \bs1,\,\bs{20},\,\bs2,\,\bs1) & {\scriptsize$(0,\,0,\,0,\,0,\,0)$}\\ 
2 & ( \bs1,\,\bs1,\,\bs1,\,\bs2)    & {\scriptsize$(0,\,0,\,0,\,0,\,\sqrt{\tfrac32})$}&
2 & ( \bs1,\,\bs1,\,\bs1,\,\bs1)    & {\scriptsize$(\sqrt{\tfrac5{14}},\,-\tfrac5{2\sqrt7},\,\tfrac1{\sqrt2},\,-\tfrac12,\,0)$}\\
2 & ( \bs1,\,\bs1,\,\bs1,\,\bs1)    & {\scriptsize$(\sqrt{\tfrac5{14}},\,-\tfrac1{\sqrt7},\,-\tfrac1{\sqrt2},\,-1,\,0)$}&
8 & ( \bs1,\,\bs6,\,\bs1,\,\bs1)    & {\scriptsize$(0,\,-\frac{\sqrt7}4,\,0,\,\frac1{4},\,-\frac1{\sqrt6})$}\\
8 & ( \bs1,\,\bs6,\,\bs1,\,\bs1)    & {\scriptsize$(0,\,0,\,-\tfrac1{\sqrt2},\,0,\,-\frac1{\sqrt6})$}&&&\\
\hline
\end{tabular}}
\vskip1mm
Consequently, the $\beta$ function coefficients are given by
$b_{\SU5}^{\maN=2}=-8,\,b_{\SU6}^{\maN=2}=b_{\SU2_a}^{\maN=2}=16$,
$b_{\SU2_b}^{\maN=2}=-2$ and
\begin{equation}
b_{\U1}^{\maN=2} =~\mbox{\scriptsize$
\left(
\begin{array}{ccccc}
\frac{19}{14}          & -\frac{3\sqrt{10}}7 & -\frac32\sqrt{\frac57} & 0 & 0\\
-\frac{3\sqrt{10}}7    & \frac{92}7          & 2\sqrt{\frac27}        &-2\sqrt7 & \sqrt{42}\\
-\frac32\sqrt{\frac57} & 2\sqrt{\frac27}     & \frac{31}2             &-\sqrt2  & 4\sqrt3\\
0                      & -2\sqrt7            & -\sqrt2                & 4       & -\sqrt{6}\\
0                      & \sqrt{42}           & 4\sqrt3                & -\sqrt{6}& 11
\end{array}
\right)\,.
$}
\end{equation}
In the notation of eq.~\eqref{eq:Neq2combination}, the overlap of the hypercharge and $\U1_X$ is given by
\begin{subequations}
\label{eqs:Z2OverlapExample1}
\begin{eqnarray}
 &\mbox{\small
  $n_Y = (-\sqrt{\tfrac2{35}},\,\tfrac1{\sqrt{7}},\,-\sqrt2,\,1,\,0),\,
  m_Y^{\SU5} = (\tfrac2{15},\,\tfrac4{15},\,\tfrac2{5},\,\tfrac1{5}),\, 
  m_Y^{b'}=0 \text{ for other }b'$}&\\
 &\mbox{\small
  $n_X = (\tfrac3{\sqrt{70}},\,\tfrac2{\sqrt{7}},\,0,\,0,\,0),\,
  m_X^{\SU5} = (-\tfrac1{5},\,-\tfrac2{5},\,-\tfrac3{5},\,-\tfrac3{10}),\, 
  m_X^{b'}=0 \text{ for other }b'$}&
\end{eqnarray}
\end{subequations}
Thus, according to eq.~\eqref{eq:mixingsimple}, $b_{YX}^2=0$.

\paragraph{\bs{\Z3} $\bs{\maN=2}$ theory.}
The unbroken gauge group is $\SU3_a\x\SU3_b\x\U1^4\x[\SU6\x\U1^3]$, and the $\maN=2$ \U1 generators are
\begin{eqnarray}
\label{eq:Z3Neq2U1sExample1}
t_{10'}&=&(1,0,0,0,0,0,0,0)(0,0,0,0,0,0,0,0)\,,\\
t_{11'}&=&(0,1,0,0,0,0,0,0)(0,0,0,0,0,0,0,0)\,,\nn\\
t_{12'}&=&(0,0,0,0,0,0,0,0)(1,0,0,0,0,0,0,0)\,,\nn\\
t_{13'}&=&(0,0,0,0,0,0,0,0)(0,1,0,0,0,0,0,0)\,,\nn\\
t_{14'}&=&\tfrac{1}{\sqrt{3}}(0,0,1,1,1,0,0,0)(0,0,0,0,0,0,0,0)\,,\nn\\
t_{15'}&=&\tfrac{1}{\sqrt{3}}(0,0,0,0,0,1,1,1)(0,0,0,0,0,0,0,0)\,,\nn\\
t_{16'}&=&\tfrac{1}{\sqrt{6}}(0,0,0,0,0,0,0,0)(0,0,1,1,1,1,1,1)\,,\nonumber
\end{eqnarray}

The $\maN=2$ matter spectrum contains the following hypermultiplet gauge representations\\
\centerline{\small
\begin{tabular}{|ccc||ccc|}
\hline
\# & Irrep & \U1 charges & \# & Irrep & \U1 charges\\
\hline
1 & ( \bs{1},\,\bs{1},\,\bs{1}) & {\scriptsize$(0,0,-1,-1,0,0,0)$} & 
1 & ( \bs{1},\,\bs{1},\,\bs{20}) & {\scriptsize$(0,0,-\frac{1}{2},\frac{1}{2},0,0,0)$}\\
1 & ( \bs{1},\,\bs{1},\,\bs{1}) & {\scriptsize$(0,0,\frac{1}{2},\frac{1}{2},0,0,-\sqrt{\frac{3}{2}})$} & 
1 & ( \bs{1},\,\bs{1},\,\bs{1}) & {\scriptsize$(0,0,\frac{1}{2},\frac{1}{2},0,0,\sqrt{\frac{3}{2}})$}\\
1 & ( \bs{1},\,\bs{1},\,\bs{1}) & {\scriptsize$(0,0,1,-1,0,0,0)$} & 
1 & ( \bs{3},\,\bsb{3},\,\bs{1}) & {\scriptsize$(\frac{1}{2},-\frac{1}{2},0,0,-\frac{1}{2 \sqrt{3}},\frac{1}{2 \sqrt{3}},0)$}\\
1 & ( \bs{1},\,\bs{1},\,\bs{1}) & {\scriptsize$(\frac{1}{2},-\frac{1}{2},0,0,\frac{\sqrt{3}}{2},-\frac{\sqrt{3}}{2},0)$} & 
1 & ( \bs{1},\,\bs{1},\,\bs{1}) & {\scriptsize$(\frac{1}{2},\frac{1}{2},0,0,-\frac{\sqrt{3}}{2},-\frac{\sqrt{3}}{2},0)$}\\
6 & ( \bs{1},\,\bs{1},\,\bs{1}) & {\scriptsize$(\frac{2}{3},0,0,\frac{2}{3},0,0,0)$} & 
3 & ( \bs{3},\,\bs{1},\,\bs{1}) & {\scriptsize$(-\frac{1}{3},0,0,\frac{2}{3},0,-\frac{1}{\sqrt{3}},0)$}\\
3 & ( \bs{1},\,\bs{1},\,\bsb{6}) & {\scriptsize$(\frac{2}{3},0,0,-\frac{1}{3},0,0,\frac{1}{\sqrt{6}})$} & 
3 & ( \bsb{3},\,\bs{1},\,\bs{1}) & {\scriptsize$(-\frac{1}{3},0,0,\frac{2}{3},0,\frac{1}{\sqrt{3}},0)$}\\
3 & ( \bs{1},\,\bs{1},\,\bs{6}) & {\scriptsize$(\frac{2}{3},0,0,-\frac{1}{3},0,0,-\frac{1}{\sqrt{6}})$}& 
3 & ( \bs{1},\,\bs{1},\,\bs{1}) & {\scriptsize$(\frac{2}{3},0,-1,-\frac{1}{3},0,0,0)$}\\
3 & ( \bs{1},\,\bs{1},\,\bs{1}) & {\scriptsize$(\frac{2}{3},0,1,-\frac{1}{3},0,0,0)$} & 
3 & ( \bs{1},\,\bs{1},\,\bs{1}) & {\scriptsize$(-\frac{5}{6},\frac{1}{2},-\frac{1}{2},\frac{1}{6},\frac{1}{2 \sqrt{3}},\frac{1}{2 \sqrt{3}},-\frac{1}{\sqrt{6}})$} \\
3 & ( \bs{1},\,\bsb{3},\,\bs{1}) & {\scriptsize$(-\frac{1}{3},0,-\frac{1}{2},\frac{1}{6},0,-\frac{1}{\sqrt{3}},-\frac{1}{\sqrt{6}})$} &
3 & ( \bs{3},\,\bs{1},\,\bs{1}) & {\scriptsize$(\frac{1}{6},\frac{1}{2},-\frac{1}{2},\frac{1}{6},\frac{1}{2 \sqrt{3}},-\frac{1}{2 \sqrt{3}},-\frac{1}{\sqrt{6}})$} \\
6 & ( \bs{1},\,\bs{1},\,\bs{1}) & {\scriptsize$(\frac{1}{6},-\frac{1}{2},-\frac{1}{2},\frac{1}{6},-\frac{1}{2 \sqrt{3}},-\frac{1}{2 \sqrt{3}},\frac{1}{\sqrt{6}})$} &
3 & ( \bs{1},\,\bs{3},\,\bs{1}) & {\scriptsize$(-\frac{1}{3},0,-\frac{1}{2},\frac{1}{6},0,\frac{1}{\sqrt{3}},\frac{1}{\sqrt{6}})$} \\
3 & ( \bs{1},\,\bs{1},\,\bsb{6}) & {\scriptsize$(\frac{1}{6},-\frac{1}{2},0,-\frac{1}{3},-\frac{1}{2 \sqrt{3}},-\frac{1}{2 \sqrt{3}},-\frac{1}{\sqrt{6}})$} &
3 & ( \bs{1},\,\bs{1},\,\bs{1}) & {\scriptsize$(\frac{1}{6},-\frac{1}{2},0,\frac{2}{3},-\frac{1}{2 \sqrt{3}},-\frac{1}{2 \sqrt{3}},-\sqrt{\frac{2}{3}})$}\\
3 & ( \bs{1},\,\bs{1},\,\bs{1}) & {\scriptsize$(\frac{1}{6},-\frac{1}{2},\frac{1}{2},-\frac{5}{6},-\frac{1}{2 \sqrt{3}},-\frac{1}{2 \sqrt{3}},\frac{1}{\sqrt{6}})$} &
3 & ( \bs{1},\,\bs{1},\,\bs{6}) & {\scriptsize$(\frac{1}{6},-\frac{1}{2},\frac{1}{2},\frac{1}{6},-\frac{1}{2 \sqrt{3}},-\frac{1}{2 \sqrt{3}},0)$} \\
3 & ( \bs{1},\,\bsb{3},\,\bs{1}) & {\scriptsize$(\frac{1}{6},\frac{1}{2},-\frac{1}{2},\frac{1}{6},\frac{1}{2 \sqrt{3}},-\frac{1}{2 \sqrt{3}},\frac{1}{\sqrt{6}})$} & &&\\
\hline
\end{tabular}
}
\vskip1mm
Consequently, the $\beta$ function coefficients are given by 
$b_{\SU3_a}^{\maN=2}=b_{\SU3_b}^{\maN=2}=b_{\SU6}^{\maN=2}=6$ and
\begin{equation}
b_{\U1}^{\maN=2} =~\mbox{\scriptsize$
\left(
\begin{array}{ccccccc}
16        & -3            &     2    & -6      & -\sqrt{3}         & -\sqrt{3} & 0 \\
-3        & 10            & -\frac92 & \frac32 & 3\sqrt3           & \sqrt3    & \sqrt{\frac32}\\
2         &-\frac92       & 15       & -4      & -\frac{3\sqrt3}{2}& 0         & 0\\
-6        & \frac32       & -4       & 15      & \frac{\sqrt3}2    & 0         & 0\\
-\sqrt{3} & 3\sqrt3       &-\frac{3\sqrt3}{2}& \frac{\sqrt3}2 &4   & 1         &\frac1{\sqrt2}\\
-\sqrt{3} & \sqrt3        & 0        & 0       & 1                 & 10        & 2\sqrt2\\
0         & \sqrt{\frac32}& 0        & 0       & \frac1{\sqrt2}    & 2\sqrt2   & 11
\end{array}
\right)\,.
$}
\end{equation}
In the notation of eq.~\eqref{eq:Neq2combination}, the overlap of the hypercharge and $\U1_X$ is given by
\begin{subequations}
\label{eqs:Z3OverlapExample1}
\begin{eqnarray}
 &n_Y=(0,0,0,0,-\tfrac2{\sqrt3},\tfrac2{\sqrt3},0),\,m_Y^{\SU3_b}=(\tfrac13,\frac23),&\quad n_Y=0,\,m_Y^{b'}=0 \text{ for other }b',\\
 &n_X=(0,1,0,0,0,0,0)\,,&\quad m_X^{b'}=0 \text{ for all }b'\,.
\end{eqnarray}
\end{subequations}
Thus, according to eq.~\eqref{eq:mixingsimple}, $b_{YX}^3=-8$.


\subsection{Example 2. Mixing in models with 3 Wilson lines}
Let us consider now the gauge embedding with 3 wilson lines:
\begin{subequations} 
\begin{eqnarray}
V &=&\tfrac1{12}(-1,-7,3,-3,-3,-3,3,3)(12,0,0,0,0,0,0,0)\;,\\
\mathcal{A}_2 &=&
\tfrac14(0,0,2,-2,-2,-2,2,2)(-3,3,3,-1,-1,-1,1,3)\;,\\ 
\mathcal{A}_2' &=&
\tfrac14(-4,-2,4,-2,-2,2,0,0)(1,3,3,-1,3,-1,-3,3)\;,\\
\mathcal{A}_3 & = &
\tfrac13(-3,3,1,-1,-1,-1,-1,-1)(1,0,0,0,0,1,-3,1)\;.
\end{eqnarray}
\end{subequations}
The interesting \U1 generators are given in the Cartan basis of \E8\x\E8 by
\begin{subequations}\label{eq:tsExample1}
\begin{eqnarray}
t_Y &=& (0,0,0,-1/2,-1/2,1/3,1/3,1/3)(0,0,0,0,0,0,0,0)\,,\\
t_X &=& \tfrac{1}{\sqrt{6}}(0,0,-1,1,1,1,1,1)(0,0,0,0,0,0,0,0)\,.
\end{eqnarray}
\end{subequations}
The matter spectrum can be obtained from~\cite{WebTables:2006xx} 
or by using the program {\small\tt{orbifolder}}~\cite{Nilles:2011aj,ORB11}.

Using the data below, the resulting kinetic mixing is
\beq
\Delta_{YX}= -\frac1{16\pi^2}\sqrt{\frac32}\log \bigg(|\eta(3\,T_2)|^4 \mathrm{Im}(T_2)\bigg)\,.
\eeq

\paragraph{\bs{\Z2} $\bs{\maN=2}$ theory.}
The unbroken gauge group is $\SU5\x\U1^4\x[\SU2\x\SU5_h\x\U1^3]$, and the $\maN=2$ \U1 generators are
\begin{eqnarray}
\label{eq:Z2Neq2U1sEx2}
t_{10'}&=&(0,0,0,0,0,0,0,0)(1,0,0,0,0,0,0,0)\,,\\
t_{11'}&=&\tfrac{1}{\sqrt{70}}(-3,7,-3,-1,-1,-1,0,0)(0,0,0,0,0,0,0,0)\,,\nn\\
t_{12'}&=&\tfrac{1}{2\sqrt{7}}(-3,0,4,-1,-1,-1,0,0)(0,0,0,0,0,0,0,0)\,,\nn\\
t_{13'}&=&\tfrac{1}{\sqrt{2}}(0,0,0,0,0,0,1,1)(0,0,0,0,0,0,0,0)\,,\nn\\
t_{14'}&=&\tfrac{1}{\sqrt{2}}(0,0,0,0,0,0,0,0)(0,0,0,0,0,1,0,1)\,,\nonumber\\
t_{15'}&=&\tfrac{1}{2}(-1,0,0,1,1,1,0,0)(0,0,0,0,0,0,0,0)\,,\nonumber\\
t_{16'}&=&\tfrac{1}{\sqrt{5}}(0,0,0,0,0,0,0,0)(0,1,1,1,1,0,-1,0)\,.\nonumber
\end{eqnarray}

The gauge representations of the $\maN=2$ matter spectrum are:\\
\centerline{\small
\begin{tabular}{|ccc||ccc|}
\hline
\# & Irrep & \U1 charges & \# & Irrep & \U1 charges\\
\hline
2 & ( {\bs{5},\, \bs{1},\, \bs{1}}) & {\scriptsize$(0,-\frac{3}{\sqrt{70}},\frac{2}{\sqrt{7}},\frac{1}{\sqrt{2}},0,0,0)$} &
2 & ( {\bs{1},\, \bs{1},\, \bs{1}}) & {\scriptsize$(0,-\sqrt{\frac{5}{14}},\frac{1}{\sqrt{7}},-\frac{1}{\sqrt{2}},0,1,0)$}\\
4 & ( {\bs{1},\, \bs{1},\, \bs{1}}) & {\scriptsize$(-1,\frac{\sqrt{\frac{5}{14}}}{2},-\frac{1}{2 \sqrt{7}},-\frac{1}{2 \sqrt{2}},0,\frac{1}{2},0)$} &
4 & ( {\bs{1},\, \bs{2},\, \bs{1}}) & {\scriptsize$(0,\frac{\sqrt{\frac{5}{14}}}{2},-\frac{1}{2 \sqrt{7}},-\frac{1}{2 \sqrt{2}},-\frac{1}{\sqrt{2}},\frac{1}{2},0)$}\\
4 & ( {\bs{1},\, \bs{1},\, \bs{1}}) & {\scriptsize$(-\frac{1}{4},\frac{\sqrt{\frac{5}{14}}}{2},\frac{5}{4 \sqrt{7}},\frac{1}{2 \sqrt{2}},\frac{1}{2 \sqrt{2}},\frac{3}{4},\frac{\sqrt{5}}{4})$}&
4 & ( {\bs{1},\, \bs{1},\, \bs{1}}) & {\scriptsize$(-\frac{1}{4},-\frac{\sqrt{\frac{5}{14}}}{2},\frac{9}{4 \sqrt{7}},-\frac{1}{2 \sqrt{2}},\frac{1}{2 \sqrt{2}},-\frac{1}{4},\frac{\sqrt{5}}{4})$}\\
4 & ( {\bs{1},\, \bs{1},\, \bs{1}}) & {\scriptsize$(\frac{1}{4},\frac{3 \sqrt{\frac{5}{14}}}{2},\frac{1}{4 \sqrt{7}},-\frac{1}{2 \sqrt{2}},-\frac{1}{2 \sqrt{2}},-\frac{1}{4},-\frac{\sqrt{5}}{4})$} &
4 & ( {\bs{1},\, \bs{1},\, \bs{1}}) & {\scriptsize$(-\frac{3}{4},\frac{\sqrt{\frac{5}{14}}}{2},\frac{5}{4 \sqrt{7}},\frac{1}{2 \sqrt{2}},-\frac{1}{2 \sqrt{2}},-\frac{1}{4},-\frac{\sqrt{5}}{4})$} \\
4 & ( {\bs{1},\, \bs{1},\, \bs{5}}) & {\scriptsize$(-\frac{1}{4},\frac{\sqrt{\frac{5}{14}}}{2},\frac{5}{4 \sqrt{7}},\frac{1}{2 \sqrt{2}},\frac{1}{2 \sqrt{2}},-\frac{1}{4},\frac{1}{4 \sqrt{5}})$} &
4 & ( {\bsb{5},\, \bs{1},\, \bs{1}}) & {\scriptsize$(0,\frac{11}{2 \sqrt{70}},\frac{1}{\sqrt{7}},\frac{1}{2 \sqrt{2}},0,0,0)$}\\
4 & ( {\bs{5},\, \bs{1},\, \bs{1}}) & {\scriptsize$(0,\frac{9}{2 \sqrt{70}},\frac{1}{2 \sqrt{7}},-\frac{1}{2 \sqrt{2}},0,\frac{1}{2},0)$} 
&&&\\
\hline
\end{tabular}}
Hence, the $\beta$ function coefficients are given by
$b_{\SU5}^{\maN=2}=b_{\SU2}^{\maN=2}=0,\,b_{\SU5_h}^{\maN=2}=-6$
and
\begin{equation}
b_{\U1}^{\maN=2} =~\mbox{\scriptsize$
\left(
\begin{array}{ccccccc}
\frac{33}{8 }             & -\frac94\sqrt{\frac5{14}}  & -\frac{45}{8\sqrt{7}}        & -\frac{5}{4\sqrt{2}}     & -\frac{5}{4\sqrt{2}}   & -\frac38                   & -\frac{\sqrt5}8 \\
-\frac94\sqrt{\frac5{14}} & \frac{165}{14}             & \frac{57}{28}\sqrt{\frac52}  & 0                        & -\frac34\sqrt{\frac57} & \frac{15}4\sqrt{\frac5{14}}& -\frac{15}{4\sqrt{14}}\\
-\frac{45}{8\sqrt{7}}     & \frac{57}{28}\sqrt{\frac52}& \frac{537}{56}               & \frac{11}4\sqrt{\frac72} & \frac{41}{4\sqrt{14}}  & -\frac9{8\sqrt7}           & \frac{13}8\sqrt{\frac57}\\
-\frac{5}{4\sqrt{2}}      & 0                          & \frac{11}4\sqrt{\frac72}     & \frac{17}2               & \frac94                & -\frac{21}{4\sqrt2}        & \frac14\sqrt{\frac52}\\
-\frac{5}{4\sqrt{2}}      & -\frac34\sqrt{\frac57}     & \frac{41}{4\sqrt{14}}        & \frac94                  & \frac{17}4             & -\frac9{4\sqrt2}           & \frac54\sqrt{\frac52}\\
-\frac38                  & \frac{15}4\sqrt{\frac5{14}}& -\frac9{8\sqrt7}             & -\frac{21}{4\sqrt2}      & -\frac9{4\sqrt2}       &  \frac{57}8                & \frac{3\sqrt5}8\\
-\frac{\sqrt5}8           & -\frac{15}{4\sqrt{14}}     & \frac{13}8\sqrt{\frac57}     & \frac14\sqrt{\frac52}    & \frac54\sqrt{\frac52}  & \frac{3\sqrt5}8            & \frac{21}8\\
\end{array}
\right)\,.
$}
\end{equation}
In the notation of eq.~\eqref{eq:Neq2combination}, the overlap of the hypercharge and $\U1_X$ is given by
\begin{subequations}
\label{eqs:Z2OverlapExample2}
\begin{eqnarray}
 &\mbox{\small
  \!\!\!\!\!\!\!\!\!\!
  $n_Y = (0,\tfrac1{\sqrt{70}},\tfrac1{2\sqrt{7}},\tfrac1{\sqrt2},0,-\tfrac12,0),\,
  m_Y^{\SU5} = (-\tfrac1{10},\,-\tfrac1{5},\,-\tfrac2{15},\,-\tfrac1{15}),\, 
  m_Y^{b'}=0 \text{ for other }b'$}&\\
 &\mbox{\small
  \!\!\!\!\!\!\!
  $n_X = (\tfrac3{\sqrt{70}},\,\tfrac2{\sqrt{7}},\,0,\,0,\,0),\,
  m_X^{\SU5} = (-\tfrac1{10},\,-\tfrac1{5},\,-\tfrac2{15},\,-\tfrac1{15}),\, 
  m_X^{b'}=0 \text{ for other }b'$}&
\end{eqnarray}
\end{subequations}
Thus, according to eq.~\eqref{eq:mixingsimple}, $b_{YX}^2=-3\sqrt{\frac32}$.

\paragraph{\bs{\Z3} $\bs{\maN=2}$ theory.}
The unbroken gauge group is $\SU3_a\x\SU3_b\x\U1^4\x[\SU2\x\SU5\x\U1^3]$, and the $\maN=2$ \U1 generators are
\begin{eqnarray}
\label{eq:Z3Neq2U1sExample2}
t_{10'}&=&(1,0,0,0,0,0,0,0)(0,0,0,0,0,0,0,0)\,,\\
t_{11'}&=&(0,1,0,0,0,0,0,0)(0,0,0,0,0,0,0,0)\,,\nn\\
t_{12'}&=&(0,0,0,0,0,0,0,0)(1,0,0,0,0,0,0,0)\,,\nn\\
t_{13'}&=&\tfrac1{\sqrt2}(0,0,0,0,0,0,0,0)(0,0,0,0,0,1,0,1)\,,\nn\\
t_{14'}&=&\tfrac1{\sqrt3}(0,0,0,1,1,1,0,0)(0,0,0,0,0,0,0,0)\,,\nn\\
t_{15'}&=&\tfrac1{\sqrt3}(0,0,-1,0,0,0,1,1)(0,0,0,0,0,0,0,0)\,,\nn\\
t_{16'}&=&\tfrac1{\sqrt5}(0,0,0,0,0,0,0,0)(0,1,1,1,1,0,-1,0)\,,\nonumber
\end{eqnarray}

The gauge representations of the $\maN=2$ hypermultiplets are \\
\centerline{\small
\begin{tabular}{|ccc||ccc|}
\hline
\# & Irrep & \U1 charges & \# & Irrep & \U1 charges\\
\hline
1 & ( {\bs{1},\, \bs{1},\, \bs{1},\, \bs{1}}) & {\scriptsize$(\frac{1}{2},\frac{1}{2},0,0,-\frac{\sqrt{3}}{2},\frac{\sqrt{3}}{2},0)$}&
1 & ( {\bs{3},\, \bs{3},\, \bs{1},\, \bs{1}}) & {\scriptsize$(\frac{1}{2},\frac{1}{2},0,0,\frac{1}{2 \sqrt{3}},-\frac{1}{2 \sqrt{3}},0)$}\\
6 & ( {\bs{1},\, \bs{1},\, \bs{1},\, \bs{1}}) & {\scriptsize$(\frac{2}{3},\frac{2}{3},0,0,0,0,0)$}&
3 & ( {\bsb{3},\, \bsb{3},\, \bs{1},\, \bs{1}}) & {\scriptsize$(\frac{1}{6},\frac{1}{6},0,0,-\frac{1}{2 \sqrt{3}},\frac{1}{2 \sqrt{3}},0)$}\\
3 & ( {\bs{1},\, \bs{1},\, \bs{1},\, \bs{1}}) & {\scriptsize$(\frac{1}{6},\frac{1}{6},0,0,\frac{\sqrt{3}}{2},-\frac{\sqrt{3}}{2},0)$}&
3 & ( {\bsb{3},\, \bs{1},\, \bs{1},\, \bs{1}}) & {\scriptsize$(\frac{1}{6},\frac{1}{6},0,0,\frac{\sqrt{3}}{2},\frac{1}{2 \sqrt{3}},0)$}\\
3 & ( {\bs{1},\, \bsb{3},\, \bs{1},\, \bs{1}}) & {\scriptsize$(\frac{1}{6},\frac{1}{6},0,0,-\frac{1}{2 \sqrt{3}},-\frac{\sqrt{3}}{2},0)$}&
3 & ( {\bs{1},\, \bs{1},\, \bs{2},\, \bs{1}}) & {\scriptsize$(-\frac{5}{6},\frac{1}{6},\frac{1}{3},-\frac{1}{3 \sqrt{2}},\frac{1}{2 \sqrt{3}},\frac{1}{2 \sqrt{3}},0)$}\\
3 & ( {\bs{1},\, \bs{1},\, \bs{1},\, \bs{1}}) & {\scriptsize$(-\frac{1}{3},-\frac{1}{3},-\frac{2}{3},\frac{\sqrt{2}}{3},-\frac{1}{\sqrt{3}},-\frac{1}{\sqrt{3}},0)$}&
3 & ( {\bs{1},\, \bs{3},\, \bs{1},\, \bs{1}}) & {\scriptsize$(\frac{1}{6},\frac{1}{6},-\frac{2}{3},\frac{\sqrt{2}}{3},-\frac{1}{2 \sqrt{3}},\frac{1}{2 \sqrt{3}},0)$}\\
3 & ( {\bs{1},\, \bs{1},\, \bs{1},\, \bs{1}}) & {\scriptsize$(-\frac{1}{3},-\frac{1}{3},\frac{2}{3},-\frac{\sqrt{2}}{3},\frac{1}{\sqrt{3}},\frac{1}{\sqrt{3}},0)$}&
3 & ( {\bs{1},\, \bs{1},\, \bs{2},\, \bs{1}}) & {\scriptsize$(\frac{1}{6},-\frac{5}{6},-\frac{1}{3},\frac{1}{3 \sqrt{2}},-\frac{1}{2 \sqrt{3}},-\frac{1}{2 \sqrt{3}},0)$}\\
3 & ( {\bs{3},\, \bs{1},\, \bs{1},\, \bs{1}}) & {\scriptsize$(\frac{1}{6},\frac{1}{6},\frac{2}{3},-\frac{\sqrt{2}}{3},-\frac{1}{2 \sqrt{3}},\frac{1}{2 \sqrt{3}},0)$}
&&&\\
\hline
\end{tabular}
}
Consequently, the $\beta$ function coefficients are given by 
$b_{\SU3_a}^{\maN=2}=b_{\SU3_b}^{\maN=2}=12,\,b_{\SU2}^{\maN=2}=2,\,b_{\SU5}^{\maN=2}=-10$ and
\begin{equation}
b_{\U1}^{\maN=2} =~\mbox{\scriptsize$
\left(
\begin{array}{ccccccc}
 6 & 3 & -1 & \frac{1}{\sqrt{2}} & -\frac{\sqrt{3}}{2} & -\frac{\sqrt{3}}{2} & 0 \\
 3 & 6 & 1 & -\frac{1}{\sqrt{2}} & \frac{\sqrt{3}}{2} & \frac{\sqrt{3}}{2} & 0 \\
 -1 & 1 & 6 & -3 \sqrt{2} & \sqrt{3} & \sqrt{3} & 0 \\
 \frac{1}{\sqrt{2}} & -\frac{1}{\sqrt{2}} & -3 \sqrt{2} & 3 & -\sqrt{\frac{3}{2}} & -\sqrt{\frac{3}{2}} & 0 \\
 -\frac{\sqrt{3}}{2} & \frac{\sqrt{3}}{2} & \sqrt{3} & -\sqrt{\frac{3}{2}} & 9 & 0 & 0 \\
 -\frac{\sqrt{3}}{2} & \frac{\sqrt{3}}{2} & \sqrt{3} & -\sqrt{\frac{3}{2}} & 0 & 9 & 0 \\
 0 & 0 & 0 & 0 & 0 & 0 & 0
\end{array}
\right)\,.
$}
\end{equation}
In the notation of eq.~\eqref{eq:Neq2combination}, the overlap of the hypercharge and $\U1_X$ is given by
\begin{subequations}
\label{eqs:Z3OverlapExample2}
\begin{eqnarray}
 &n_a=\tfrac1{\sqrt3}(0,0,0,0,-1,1,0),\,m_a^{\SU3_b}=(\tfrac16,\frac13),&\quad n_a=0,\,m_a^{b'}=0 \text{ for other }b',\\
 &n_b=(0,0,0,0,\tfrac1{\sqrt2},\tfrac1{\sqrt2},0)\,,&\quad m_b^{b'}=0 \text{ for all }b'\,.
\end{eqnarray}
\end{subequations}
Thus, according to eq.~\eqref{eq:mixingsimple}, $b_{YX}^3=0$.

\providecommand{\newblock}{}

\end{document}